\documentclass[showpacs,preprintnumbers,twocolumn,amsmath,amssymb,
superscriptaddress,floatfix]{revtex4}

\usepackage{graphicx}

\begin{document}
\title{Nonuniversal prefactors in 
correlation functions of 1D quantum liquids}

\author{Aditya Shashi}
\affiliation{Department of Physics and Astronomy, Rice
University, Houston, Texas 77005, USA}
\author{Leonid~I.~Glazman}
\affiliation{Department of Physics, Yale University, 217 Prospect
Street, New Haven, Connecticut 06520, USA}
\author{Jean-S{\'e}bastien Caux}
\affiliation{Institute for Theoretical Physics, Universiteit van
Amsterdam, 1090 GL Amsterdam, The Netherlands}
\author{Adilet~Imambekov}
\affiliation{Department of Physics and Astronomy, Rice
University, Houston, Texas 77005, USA}
\date{\today}

\newcommand{\tpsi}{\tilde{\psi}}
\newcommand{\tV}{\tilde{V}}
\newcommand{\hpsi}{\hat{\psi}}
\renewcommand{\d}{\hat{d}}
\newcommand{\trho}{\tilde{\rho}}
\newcommand{\hrho}{\hat{\rho}}
\newcommand{\alphabar}{{\bar{\alpha}}}
\newcommand{\expct}[1]{\left\langle #1 \right\rangle}
\newcommand{\expcts}[1]{\langle #1 \rangle}
\renewcommand{\mod}{\textrm{mod }}
\newcommand{\etal}{\textit{et al.}}
\renewcommand{\th}{{\rm th}}
\newcommand{\bw}{\begin{widetext}}
\newcommand{\ew}{\end{widetext}}
\newcommand{\be}{\begin{equation}}
\newcommand{\ee}{\end{equation}}
\newcommand{\bea}{\begin{eqnarray}}
\newcommand{\eea}{\end{eqnarray}}

\newcommand{\ceq}[1]{Eq.~(\ref{#1})}
\newcommand{\dg}{^{\dagger}}
\newcommand{\p}{\hat{\psi}}
\newcommand{\V}{\hat{V}}
\newcommand{\al}{|\alpha\rangle}
\newcommand{\pd}{\hat{\psi}\dg}

\begin{abstract}
We develop a general approach to calculating ``nonuniversal" prefactors in static and dynamic correlation functions of 1D quantum liquids at zero temperature,
by relating them to the finite size scaling of certain matrix elements (form factors). This represents a new, powerful tool for extracting data valid in the thermodynamic limit from finite-size effects.
As the main application, we consider weakly interacting spinless fermions with an arbitrary pair interaction potential, for which 
we perturbatively calculate certain prefactors in static and dynamic correlation functions.
We also non-perturbatively evaluate prefactors  of the long-distance behavior of correlation functions for the exactly solvable Lieb-Liniger model of 1D bosons.

\end{abstract}


\maketitle

\section{Introduction}

One-dimensional (1D) quantum liquids of bosons, fermions and spins are conventionally described using an effective hydrodynamic approach known as the Luttinger liquid theory~\cite{EL,Haldane,Caza04,Gbook, GNT}. This theory predicts the long-range behavior of equal-time correlation functions at zero temperature, which one obtains as a series expansion with power laws controlled by a dimensionless Luttinger liquid parameter $K>0,$ see Eqs.~(\ref{Amdef})-(\ref{Cmdef}). While the ``universal'' parameter $K$ is related to thermodynamic properties and can be easily extracted from numerical or exact solutions, the ``nonuniversal'' prefactors in the series expansion, e.g. $A_m, B_m, C_m$, see Eqs.~(\ref{Amdef})-(\ref{Cmdef}) are usually not known except for a few cases~\cite{Popov_prefactor,K1,lukyanov03,CS_results,Kitanine_review}. At the same time, these prefactors set the actual scale of observable correlations, consequently determining them is an important theoretical challenge.

In this article, we first develop a general technique for calculating these nonuniversal prefactors by combining the Luttinger liquid Hamiltonian with the analysis of the finite-size properties of certain matrix elements (form factors).
We then consider dynamic response functions such as the density structure factor and the spectral function, see Fig.~\ref{Fig_def}. It has been shown recently~\cite{Pustilnik2006Fermions,Pustilnik_solo_PRL,Khodas2006Fermions,PRL_08, Zvonarev, universal,PRL_09,PRL_10,XXZ} that dynamic response functions generically have singularities which can be described by effective Hamiltonians of impurities moving in Luttinger liquids.
Analysis of the finite-size properties of these effective Hamiltonians allows us to extend the approach to various dynamic response functions. To demonstrate it, we perturbatively evaluate several prefactors of equal-time correlation functions and dynamic response functions for weakly-interacting fermions.  We also calculate, non-perturbatively, various prefactors for the exactly solvable Lieb-Liniger model~\cite{LL, Korepin} of 1D bosons with contact interactions. The latter model has been realized with ultracold atomic gases~\cite{dweiss}, and its correlation functions can be measured using
interference~\cite{interference, exp2}, analysis of particle losses~\cite{3decay}, photoassociation~\cite{3decay},
or Bragg and photoemission
spectroscopy~\cite{BraggPE}.

This article is organized as follows.
In Sec. IIA, we use linear Luttinger liquid theory to work out the connection 
between prefactors of equal-time correlation functions and lowest  energy form factors. In Sec. IIB, 
we show that the relative spectral weights of all low energy form factors 
can be fixed based on universal nonlinear Luttinger liquid theory~\cite{universal}. In Sec IIC, we show that the effective field theory of impurities moving in Luttinger liquids allows to
extend the relations between form factors and prefactors to dynamic response functions.  In Sec. III we present the results of the perturbative calculations of various prefactors for weakly interacting spinless fermions. We summarize our results in Sec. IV. Some of the technical details are contained in the Appendices.

\section{Results from effective field theory}
\subsection{Prefactors of equal-time correlators 
from the Luttinger liquid theory}

The Luttinger liquid theory~\cite{EL,Haldane,Caza04,Gbook, GNT} predicts the behavior of the correlation functions for spinless bosons and fermions of density $\rho_0$ at $\rho_0
x\gg 1$ as (here $k_F=\pi \rho_0$)
\bea
 \frac{\langle \hat \rho(x) \hat \rho(0)\rangle }{\rho_0^2} \approx&&\!\!\!\!\!\!\!\!\!\! 1\!-\! \frac{K}{2(\pi \rho_0 x)^2}+\!\!\sum_{m\geq 1}
\frac{A_m \cos(2 m k_F x)}{\left(\rho_0x\right)^{2m^2 K}}, \label{Amdef} \\
\frac{\langle \hat \psi_B^{\dagger}(x) \hat \psi_B(0) \rangle}{\rho_0} \!\!&\approx&\!\! \sum_{m\geq 0} \frac{B_m \cos(2m k_F x)}{\left(\rho_0x\right)^{2m^2
K+1/(2K)}},\label{Bndef}\\
\frac{\langle \hat \psi_F^{\dagger}(x) \hat \psi_F(0) \rangle}{\rho_0} \!\!&\approx&\!\!\sum_{m\geq 0} \frac{C_m \sin{[(2m+1)k_F x]}}{(\rho_0 x)^{(2m+1)^2 K/2+1/(2K)}}.
\label{Cmdef}
\eea
Here $\hat \rho$ is the density operator, and  $\hat \psi_F (\hat \psi_B)$ is the fermionic (bosonic) annihilation operator. The Hamiltonian describing these correlations is written as (we follow notations of Ref.~\cite{Gbook})
\bea
H_0= \frac{v}{2\pi}\int dx\;\left( K (\nabla \theta)^2 +\frac1{K}(\nabla \phi)^2\right ), \label{hbosonic}
\eea
where $v$ is the sound velocity, the canonically conjugate fields $\phi(x),\theta(x)$ have the
commutation relation $ [\phi(x),\nabla \theta(x')]=i \pi \delta(x -x'),$ and the components of the fermionic (bosonic) fields with momenta $(2m+1/2\pm 1/2)k_F$ are written as
\bea
 \psi_{F(B)}(x,t)\sim \sum_m e^{i(2m+1/2\pm 1/2)[k_Fx -\phi(x,t)]+i\theta(x,t)}.
 \eea
  For repulsive bosons, one has $K>1,$ while for repulsive (attractive) fermions $K<1(>1).$

One of the reasons for the success of the Luttinger liquid theory is its ability to predict certain finite-size effects~\cite{AffleckCardy,Caza04} due to the conformal invariance of the Hamiltonian~(\ref{hbosonic}). Below we will show that conformal invariance can also be used to relate nonuniversal prefactors to the scaling of certain form factors [see Eqs.~(\ref{fermion_scaling})-(\ref{density_scaling}), (\ref{Ascaling})], which constitutes the main result of this article. Form factors can be evaluated perturbatively or numerically for finite-size systems, and are known for certain integrable models such as the XXZ~\cite{Kitanine_JMP_09, Kitanine}, the Calogero-Sutherland~\cite{CSMexact}, and the Lieb-Liniger~\cite{Korepin, Slavnov} models. Thus relations~(\ref{fermion_scaling})-(\ref{density_scaling}),(\ref{Ascaling}) provide a powerful tool with which one may interpret finite-size effects and  make predictions which are valid in the thermodynamic limit.

Let us start by considering interacting spinless fermions. Using the resolution of the identity in the expectation value $\langle \hat \psi_F^{\dagger}(x,t)\hat \psi_F(0)\rangle,$
we get
\bea
\langle \hat \psi_F^{\dagger}(x,t)\hat \psi_F(0)\rangle=\sum_{k,\omega}e^{i (kx-\omega t)}\left|\langle k ,\omega|\hat \psi_F|N\rangle\right|^2, \label{Lehmann}
\eea
where $\langle k ,\omega  |\hat \psi_F|N\rangle$ is a form factor of the annihilation operator, $|k ,\omega \rangle$ denotes  an eigenstate of $N-1$ particles with momentum $k$ and energy $\omega$, and $|N\rangle$ is the ground state of $N$ particles. For simplicity we assume $N$ is odd so the ground state is non-degenerate. For a finite system, $k$  and $\omega$ are not continuous, but will be quantized and consequently the spectral function is a collection of delta functions in $(k,\omega).$ We will now obtain a similar representation from the Luttinger liquid theory and compare it with Eq.~(\ref{Lehmann}) to obtain the nonuniversal prefactors $C_m.$ Hamiltonian (\ref{hbosonic})  can be written using left- and right-moving components $\varphi_{L(R)}=\theta \sqrt{K} \pm \varphi/\sqrt{K},$ ~\cite{Caza04} which dictates the time dependence of the $e^{i(2m+1)k_Fx}$ component of $\langle \hat \psi^{\dagger}_F(x,t) \hat \psi_F(0,0)\rangle/\rho_0$  at $\rho_0 |x\pm vt| \gg 1$ as
\bea
\frac{e^{i(2m+1)k_Fx}}{2i(-1)^{m}}\frac{C_m \rho_0^{-(2m+1)^2K/2-1/2K}}{\left(i
(vt+x)+0\right)^{\mu_{F,L}}\left(i
(vt-x)+0\right)^{\mu_{F,R}}},\label{tdep}
\eea
where
\bea
\mu_{F,L(R)}=(2m+1)^2K/4\pm (2m+1)/2+1/4K\geq0. \label{muFdef}
\eea
 The coefficients $C_m$ appeared in \ceq{tdep} due to the comparison of the $t=0$ limit of $\langle \hat{\psi}^\dagger_F(x,t)\hat{\psi}_F(0,0)\rangle/\rho_0$ with the right hand side of \ceq{Cmdef}. The two factors in the denominator of Eq.~(\ref{tdep})
describe contributions from left (right)-going excitations near the Fermi points which
propagate with velocities $\mp v$. The signs of the infinitesimal
shifts in the denominators ensure that only excitations with
negative (positive) momenta can be created at the respective
branches. For a finite system
with periodic boundary conditions on a circle of length $L,$ conformal invariance dictates (see e.g. Ref.~\cite{Caza04}) that Eq.~(\ref{tdep}) gets modified as
\bea
\frac{e^{i(2m+1)k_Fx}C_m}{2i(-1)^{m}} \prod_{L,R}\left(\frac{\pi
e^{i\pi(vt\pm x)/L}}{i\rho_0L\sin{\frac{\pi(vt\pm x)}{L}}+0}\right)^{\mu_{F,L(R)}}. \label{eq:Cazalilla}
\eea
We can now  expand the terms in the parentheses in a Fourier series
\bea
\left(\frac{\pi
e^{i\pi(vt\pm x)/L}}{iL\sin{\frac{\pi(vt\pm x)}{L}}+0}\right)^{\mu}\!\!=\sum_{n_\mp\geq
0}  C(n_\pm,\mu)\frac{e^{\pm 2i\pi
n_{\mp}\frac{x\mp vt}{L}}}{(L/2\pi)^{\mu}},\nonumber \\
\label{eq:finite_size}
C(n_\pm,
\mu)=\frac{\Gamma(\mu+n_\pm)}{\Gamma(\mu)\Gamma(n_\pm+1)}.\ \ \ \ \ \ \ \ \ \ \ \ \ \ \  \ \ \ \
\eea
Comparing Eqs.~(\ref{eq:Cazalilla})-(\ref{eq:finite_size}) to the right hand side of Eq.~(\ref{Lehmann}),  one can clearly identify
contributions from low-energy and momenta particle-hole excitations at the right (left)
Fermi branches with energies $2\pi vn_{\pm}/L>0$ and
momenta $\pm 2\pi n_{\pm}/L$. Additionally, $m$ inter-branch pairs of momentum $2k_F$ each are formed by successively depleting discrete states below the left Fermi point $(m>0)$ and occupying the lowest-available states above the right one. On top of that, an additional hole is formed on the left branch, giving a total contribution of $(2m+1)k_F>0$ to the momentum. The contribution from $n_+=n_-=0$ gives the scaling of the ``parent'' form factor
\bea
\left|\langle m, N-1|\hat \psi_F|N\rangle\right|^2 \approx \frac{ C_m \rho_0}{2(-1)^{m}}  \left(\frac{2\pi}{\rho_0 L}\right)^{\frac{(2m+1)^2 K^2+1}{2K}},\label{fermion_scaling}
\eea
where $|m,N-1\rangle$ is the lowest energy state of $N-1$ fermions with momentum $(2m+1)k_F.$ For Galilean invariant systems, states with different $m$ can be obtained from the ground state by a center of mass Galilean boost. We see that as a consequence of the criticality of the Luttinger liquid,
form factors of the annihilation operator have nontrivial scaling with the system size, and the prefactors in front of these nontrivial powers of $L$ are directly related to the prefactors
in the correlation functions. For density correlations and bosons, similar relations can be worked out, and are given by
\bea
\left|\langle m, N-1|\hat \psi_B|N\rangle\right|^2 &\approx& \frac{B_m \rho_0 (-1)^{m}}{(2 - \delta_{0,m})}  \left(\frac{2\pi}{\rho_0 L}\right)^{\frac{4m^2 K^2+1}{2K}}, \ \ \ \ \ \!\label{boson_scaling}\\
\left|\langle m, N|\hat \rho |N\rangle\right|^2 &\approx& \frac{A_m \rho_0}{2}  \left(\frac{2\pi}{\rho_0 L}\right)^{2m^2 K}\label{density_scaling}.
\eea
Eqs.~(\ref{fermion_scaling})-(\ref{density_scaling}) allow one to evaluate the prefactors in Eqs.~(\ref{Amdef})-(\ref{Cmdef}) by identifying a single, simplest ``parent" form-factor for each of the operators $\hat{\rho},\hat{\psi}_B$ and $\hat{\psi}_F$, respectively.
 Results for bosons are simply generalized to describe
Luttinger liquids of spins  on a lattice with standard substitutions~\cite{Gbook}, and in particular
Eq.~(\ref{density_scaling}) explains the coincidence noticed in Refs.~\cite{Kitanine_review,Kitanine_JMP_09} for the spin-$1/2$ XXZ model.

\begin{figure}
\includegraphics[width=8.5 cm,height= 7.36 cm]{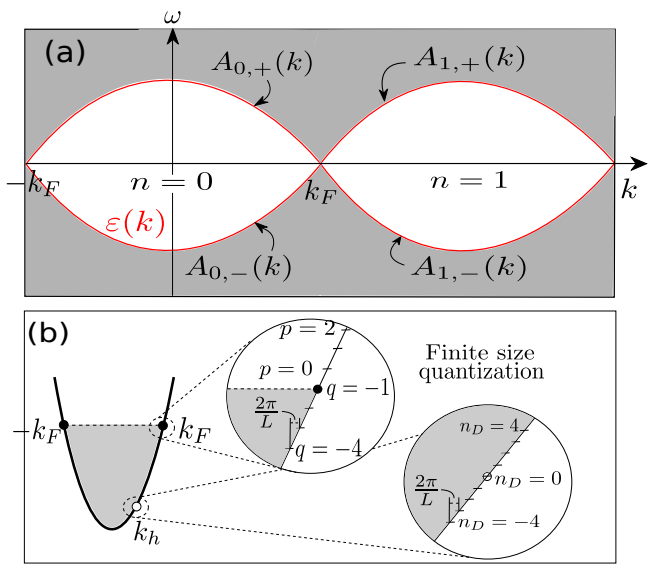}
\caption{\label{Fig_def} (Color online) (a) Spectral function $A(k,\omega),$ with shaded areas indicating the regions where it is nonzero, and notations for prefactors. (b)  ``Parent" state responsible for the singularity at $\omega \approx -\varepsilon(k)>0$ in spectral function: it contains a hole corresponding to a mobile ``impurity" at $k_h\approx-k,$ and one particle at each Fermi point. We also illustrate the finite size quantization of the momenta of the impurity  and of excitations at the right Fermi point.}
\end{figure}

\subsection{Distribution of spectral weights among low energy form factors and the multiplet summation rule}

Let us now consider the spectral weights at finite energies and momenta in the vicinity of the Fermi points (for concreteness we will discuss fermions). In a finite-size system, the spectral function is a collection of delta functions in $(k,\omega)$, weighted by form factors, see Eq.~(\ref{Lehmann}). From field theory, we not only determine the parent form factor, but also form factors associated with states containing low-energy excitations on top of the parent state, described by nonzero $n_+$ and $n_-.$Within the linear spectrum approximation
near the Fermi points,  for $n_{\pm}>1$ certain particle-hole states are degenerate, thus within $\propto 1/L$ accuracy, from Eq.~(\ref{Lehmann}) and Eqs.~(\ref{eq:Cazalilla})-(\ref{eq:finite_size}) the total spectral weight which falls into the degenerate subspace (``multiplet'') with given $n_+,n_-$ can be obtained from the parent form factor
by multiplying it by $C(n_+, \mu_{F,R})C(n_-, \mu_{F,L}).$ However, within $\propto 1/L^2$  accuracy, states with sufficiently
large $n_+$ and $n_-$ are not degenerate due to the nonlinearity of the generic fermionic spectrum, as can be illustrated by the case of weakly interacting fermions. Then $C(n_+, \mu_{F,R})C(n_-, \mu_{F,L})$ gives only the total spectral weight within each multiplet, and the conventional linear Luttinger liquid theory doesn't distinguish the $\propto 1/L^2$ splitting
of the contributions within each multiplet. However, the splitting of the spectral weights within each multiplet is also universal, and can be understood based on the universal theory of
nonlinear Luttinger liquids developed recently~\cite{universal}. Below we will illustrate such splitting for the case of $n_-=0,$ i.e. when only excitations at the right branch are created.

  Within the nonlinear Luttinger liquid theory, we evaluate various correlators (e.g. Eq.~(\ref{Amdef}),(\ref{Cmdef})) by first expressing the fermionic creation and annihilation operators in terms of fermionic quasiparticle operators $\tilde \psi_{R(L)}$. The fermionic operators are related to the quasi particle operators using
  \be
  \psi_R^\dagger (x) = e^{i\int_0^x dy (\delta_+ \tilde \psi^\dagger_R(y)\tilde \psi_R(y) + \delta_- \tilde\psi^\dagger_L(y)\tilde\psi_L(y))} \tilde\psi^\dagger_R(x).
  \ee
 Taking the nonlinearity in the spectrum of the original fermions into account~\cite{universal}, we obtain the Hamiltonian  $H_{R}+H_L$ in terms of the quasiparticle operators with
 \bea
 H_{s=R,L}=\int dx \left(\mp iv:\tilde \psi^{\dagger}_{s}\nabla \tilde \psi_{s}:+ \frac1{2m^*} :\nabla \tilde \psi^{\dagger}_{s} \nabla \tilde \psi_{s}:\right), \nonumber\\\label{HR}
 \eea
 where $m^*$ is the effective mass, characterizing the nonlinearity of $\varepsilon(k)$. 
Its inverse can be expressed through low energy parameters as \cite{Periera_PRL_06}
\bea
\frac{1}{m^*} = \frac{\partial^2\varepsilon}{\partial k^2}\bigg|_{k_F} = \frac{v}{\sqrt{K}}\frac{\partial v}{\partial h} + \frac{v^2}{2K\sqrt{K}}\frac{\partial K}{\partial h}, 
\eea
 with $h$ denoting chemical potential and $v$ the sound velocity. Expressions for correlators will include contributions from the left and right Fermi point with given $\mu_{F,R(L)}$, i.e. terms
\bea
\label{correlator}
&\propto&\bigg  \langle {\rm exp}\left[- 2\pi i \sqrt{\mu_{F,R}}\int_{-\infty}^{x}{\rm d}y :\tilde \psi^{\dagger}_R(y,t)\tilde \psi_R(y,t):\right]\times\nonumber\\
& &{\rm exp}\left[2\pi i\sqrt{\mu_{F,R}} \int_{-\infty}^{0}{\rm d}y:\tilde \psi^{\dagger}_R(y,0)\tilde \psi_R(y,0):\right]\bigg \rangle_{H_{R}}
\eea
 where the average is taken over the infinite chiral Fermi sea $|{\rm FS}\rangle$ with all negative momenta occupied, and normal ordering is with respect to this vacuum. A similar contribution from the left-movers will also appear.
 
 In a finite system, momenta of the fermionic quasiparticles are quantized near the Fermi point in increments of $2\pi/L$. We denote by $p_1 > p_2 >....>p_n\geq 0$ particle excitations carrying momenta $2\pi p_i/L$, and by $q_1<q_2<...<q_n<0$ hole excitations carrying momenta $-2\pi q_i/L$. Then $2\pi n_{+}/L = 2\pi/L\sum_i p_i - q_i $ is the total momentum  of particle-hole excitations near the right  Fermi point. Within the linear spectrum approximation, all states with the same $n_+$ are degenerate, while due to nonlinearity each of them acquires an
 energy shift 
 \bea
 \varepsilon(\{p_i,q_i\})=\frac{2\pi^2}{m^*L^2}\sum_i\left(p_i^2+q_i^2\right).
 \eea

 We can introduce a complete set of intermediate  states in \ceq{correlator} between the two exponents and organize them by the momenta of particle-hole excitations as follows
 \bea
 & &\sum_{n_+=0}^{\infty}\sum_{p_i - q_i = n_+} e^{\frac{2\pi i n^+}{L}(x-vt)-i\varepsilon(\{p_i,q_i\})t}\nonumber\\
 & & \bigg|\left \langle \{p_i,q_i\}\left|e^{2\pi i \sqrt{\mu_{F,R}} \int_{-\infty}^{0}{\rm d}y :\tilde \psi^\dagger_R(y,0)\tilde \psi_R(y,0):}\right|{\rm FS}\right \rangle\bigg|^2\nonumber\\
&=& \sum_{n_+=0}^{\infty}\sum_{p_i - q_i = n_+} e^{\frac{2\pi i n^+}{L}(x-vt)-i\varepsilon(\{p_i,q_i\})t}\nonumber\\
& & \bigg|\left \langle \{p_i,q_i\}\left|e^{-\sum_{k\neq0,l}\frac{\sqrt{\mu_{F,R}}}{k} \tilde \psi^\dagger_R(k+l)\tilde \psi_R(l)}\bigg|{\rm FS}\right \rangle\right|^2, \label{stringcorr}
\eea
where in the first line we have moved the position and time dependence of the operators over to the states, and in the second line we have Fourier transformed the creation and annihilation operators and performed the integral over $y$. The correlator above can be exactly evaluated using the methods of Ref.~\cite{BAW06}. There they consider a ``boundary state'' ${\rm exp}\left[-(a+m)\sum_{k\neq 0} \left(\frac{e^{ikx}}{k}\sum_p \pd_{p}\p_{p+k}\right)\right]|0\rangle$, which we can identify as the one obtained by action of the string operator in Eq.~(\ref{stringcorr}) on the infinite chiral vacuum if we map $ m+a \to \sqrt{\mu_{F,R}}$. Consequently we obtain the following result from Eqs.(59)-(61) of Ref.~\cite{BAW06}:

\bea
\label{f_def}
& &f(\{p_i,q_i\}) \nonumber\\
&=&\frac{\left \langle \{p_i,q_i\}\left|e^{2\pi i \sqrt{\mu_{F,R}} \int_{-\infty}^{0}{\rm d}y :\tilde \psi^\dagger_R(y,0)\tilde \psi_R(y,0):}\right|{\rm FS}\right \rangle}{\left \langle {\rm FS} \left|e^{2\pi i \sqrt{\mu_{F,R}} \int_{-\infty}^{0}{\rm d}y :\tilde \psi^\dagger_R(y,0)\tilde \psi_R(y,0):}\right|{\rm FS}\right \rangle}\nonumber\\
&=& {\rm Det}_{i,j \leq n}\left(\frac{1}{p_i-q_j}\right)\prod_{i\leq n} f^+(p_i)f^-(q_i), 
\eea
where
\bea
& &f^+(p)=\frac{\Gamma(p+1 - \sqrt{\mu_{F,R}})}{\Gamma(-\sqrt{\mu_{F,R}})\Gamma(p+1)},\ \nonumber\\
& & f^-(q) = \frac{\Gamma(-q+\sqrt{\mu_{F,R}})}{\Gamma(1+\sqrt{\mu_{F,R}})\Gamma(-q)}.
\eea

Normalization of the spectral weight leads to the following ``multiplet summation rule" (see Appendix A for an explicit demonstration)
\bea
\sum_{\sum p_i-q_i=n_+}\left|f(\{p_i,q_i\})\right|^2= C(n_+,\mu_{F,R}).
\label{mrule}
\eea

Contributions from the left Fermi point are accounted for similarly, and the total form factor is a product of these two terms.

\subsection{Prefactors of dynamic response functions from effective theory of impurities moving in Luttinger liquids}
We now consider the dynamic response functions: the density structure factor
\bea
S(k,\omega) = \int dx dt^{i(\omega t - k x)} \langle \hat{\rho}(x,t)\hat{\rho}(0,0)\rangle,\label{DSF}
\eea
and spectral function $A(k,\omega) = -\frac{1}{\pi}{\rm Im} G(k,\omega){\rm sign} \omega$ where the Green's function $G(k,\omega)$ is defined as~\cite{AGDbook}
\bea
G(k,\omega) = -i\int dx dt e^{i(\omega t - kx)} \langle T[\hat{\psi}(x,t)\hat{\psi}^{\dagger}(0,0)]\rangle.\label{Spectral}
\eea

To be specific let us consider the spectral function for $\omega>0$ and $-k_F<k<k_F.$
In addition to the Fermi points, the field theoretical description of the singularity at $-\varepsilon(k)>0$ (see Ref.~\cite{PRL_09}; we follow the notations contained therein)
involves a mobile ``impurity'' with momentum $k_h\approx-k$ moving with velocity
$v_d=\partial\varepsilon(k_h)/\partial k_h,$ see Fig.~\ref{Fig_def}. For non-interacting fermions, any spectral weight is absent
at $\omega>0$ and $-k_F<k<k_F,$ and for weakly interacting fermions the configuration responsible for a feature at $\omega\approx -\varepsilon(k)$
is illustrated in Fig.~\ref{Fig_def}; it has one particle at each Fermi points, and a hole corresponding to the impurity. While for stronger interactions such a simple interpretation
of the impurity is absent, the field theoretical description still remains valid~\cite{PRL_09}. The Hamiltonian used in this approach takes the form
\bea
H_d &=& \int dx d^{\dagger}(x)\left[ \varepsilon(k) - iv_d \frac{\partial}{\partial x}\right] d(x),\nonumber\\
H_{int} &=& \int dx \left[ V_R \rho_R(x) + V_L\rho_L(x)\right]\rho_d(x)\nonumber\\
&=& \int dx \left(V_R\nabla\frac{\theta - \phi}{2\pi} - V_L\nabla\frac{\theta + \phi}{2\pi}\right)d(x)d^{\dagger}(x).\nonumber\\
\eea
Here the operator $d(x)$ creates a mobile hole with momentum $k$ and velocity $v_d = \partial\varepsilon(k)/\partial k$. The interaction Hamiltonian describes the impurity interacting with the left and right movers of the Luttinger liquid. 
 
 The spectral function $A(k,\omega)$ in the vicinity of $-\varepsilon(k)$ is
written as
\bea
A(k,\omega)&\propto& \int dx dt e^{i\omega t}\langle d^{\dagger} (x,t)d(0,0)\rangle_{H_{LL} + H_d + H_{int}}\nonumber\\
&=& A_{0,+}(k)\int dx dt e^{i\delta \omega t}D(x,t)L(x,t)R(x,t),\nonumber\\ \label{Aint}
\eea
where $\delta\omega=\omega+\varepsilon(k),
D(x,t)=\delta(x-v_d t)$ is the impurity correlator,
$L(R)(x,t)=(i(vt \pm x)+0)^{-\mu_{0,+,L(R)}}$~\cite{muRLdef}, and we
introduced a prefactor $A_{0,+}(k)$ which will be determined by a
comparison to the form factors. After the $x,t$ integration, Eq.~(\ref{Aint}) results in
\bea
A(k,\omega)= \theta(\delta\omega ) \frac{2\pi A_{0,+}(k)
\delta\omega^{-\mu_{0,+}} }{\Gamma(1-\mu_{0,+})(v+v_d)^{\mu_{0,+,L}}(v-v_d)^{\mu_{0,+,R}}}.\nonumber
\eea
In finite-size systems, $L(x,t)$ and $R(x,t)$ get modified, see Eq.~(\ref{eq:finite_size}). The change of $D(x,t)$ to  $\sum_{n_D} e^{2i\pi
n_D(x-v_dt)/L}$ corresponds to the quantization of the impurity momentum.
 At fixed $k,$ the shift of the momentum of the impurity can be expressed
as $n_D=n_--n_+.$ Combining these terms, we get
 \bea
 & &A(k,\omega)=\nonumber\\
 & &\sum_{n_\pm \geq 0}
 \delta\left(\delta\omega-\Delta E - \frac{2\pi n_+}{L}(v-v_d)-\frac{2\pi
 n_-}{L}(v+v_d)\right) \times \nonumber \\
& &A_{0,+}(k) \frac{(2\pi)^{2-\mu_{0,+}}}{L^{1-\mu_{0,+}}} C(n_+,\mu_{0,+,R}) C(n_-,\mu_{0,+,L}),
 \label{eq:fsize}
\eea
where $\Delta E$ is a universal $\propto 1/L$ shift of the edge position~\cite{XXZ}. 
Thus the finite size structure of the response function is
given by the sum of two generically incommensurate  frequency ``ladders'' at arbitrary $k$, in contrast to the vicinities
of Fermi points.
Analysis of the scaling of the parent form factor with
$n_+=n_-=0$ then leads to
\be
\left|\langle k;N+1|\hat \psi^{\dagger}_F|N\rangle\right|^2\approx A_{0,+}(k)\left(\frac{2\pi}{L}\right)^{2-\mu_{0,+}}, \label{Ascaling}
\ee
where $|k;N+1\rangle$ denotes the lowest energy state of $N+1$ fermions with total (discrete) momentum $k.$
Similar relations can be derived for the density structure factor and the boson spectral function, and for each $k$, the left hand side consists of a single form factor which connects the ground state with the lowest energy state at total momentum $k,$ while the right hand side shows scaling with the exponents of Ref.~\cite{PRL_09}:

\bea
\left|\langle k;N|\hat\rho|0,N\rangle\right|^2&\approx& \frac{S_{0}(k)}{L}\left(\frac{2\pi}{L}\right)^{1-\mu_{0}}, \label{DSFscaling}\\
\left|\langle k;N+1|\hat\psi^{\dagger}_B|N\rangle \right|^2 &\approx& A^{B}_{0,+}(k)\left(\frac{2\pi}{L}\right)^{2-\mu^{b}_{0,+}},\label{bAscaling}
\eea
with \cite{PRL_09, muRLdef}
\bea
\label{mueqs}
\mu_{0,+} &=& 1 - \mu_{0,+,L} - \mu_{0,+,R},\nonumber\\
\mu^{b}_{0,+} &=& 1 - \mu^b_{0,+,L} - \mu^b_{0,+,R},\nonumber\\
\mu_{0} &=&1 - \mu_{0,L} - \mu_{0,R}.
\eea

 We note that in Eqs.~(\ref{Ascaling})-(\ref{bAscaling}) $k$ has to be fixed before taking the limit $L\rightarrow \infty,$ since e.g. the $k\rightarrow k_F$ and $L\rightarrow \infty$ limits do not commute as has been shown in nonlinear Luttinger liquid theory~\cite{PRL_08,universal,PRL_09}.

Eqs.~(\ref{fermion_scaling})-(\ref{density_scaling}),(\ref{Ascaling})-(\ref{bAscaling}) rely on the structure of the low-energy excitations at a given momentum, prescribed by the field theory; they are valid for all Luttinger liquids in 1D irrespective of microscopic interactions and can be used as a convenient tool to interpret the results of numerical studies. Below we illustrate their power by obtaining new nontrivial results for weakly interacting fermions and also present numerical data on some exact results ~\cite{Numerical_prefactors} we obtained from the analysis of the finite size form factors of the Lieb-Liniger model.

\section{Perturbative calculation of form factors for weakly interacting spinless fermions}

We use the non interacting Fermi gas in 1D as our unperturbed state. The ground state for a system of $N$ non interacting fermions of mass $M$ occupying a length $L$ with density $\rho_0 = N/L$ is characterized in momentum space by $N$ real momenta $\{-k_F,..., k_F\}$, increasing by increments of $2\pi/L$, with $k_F = \pi(N-1)/L$. Here we have assumed that $N$ is odd so that the ground state is non-degenerate. 

Defining the ground state $|{\rm FS}\rangle$ to have resulted from the action of $N$ creation operators on an empty vacuum starting from the left-most momentum gives us a convention to specify the relative phases of various states that will be used in the calculations to follow. Moreover in the subsequent calculations it is only the relative phases between states that is pertinent since we are interested in absolute squared values of the form factors.

To this system we add a weak four fermion interaction
\be
\label{interaction}
\V = \frac{1}{2L}\sum_{q,p,p'} V(q)\pd_{p+q}\pd_{p'-q}\p_{p'}\p_{p},\nonumber
\ee
where $V(q)$ is the Fourier transform of the pair interaction potential $V(r)$. 

\begin{figure}
\includegraphics[width=8.5 cm,height=7.5cm]{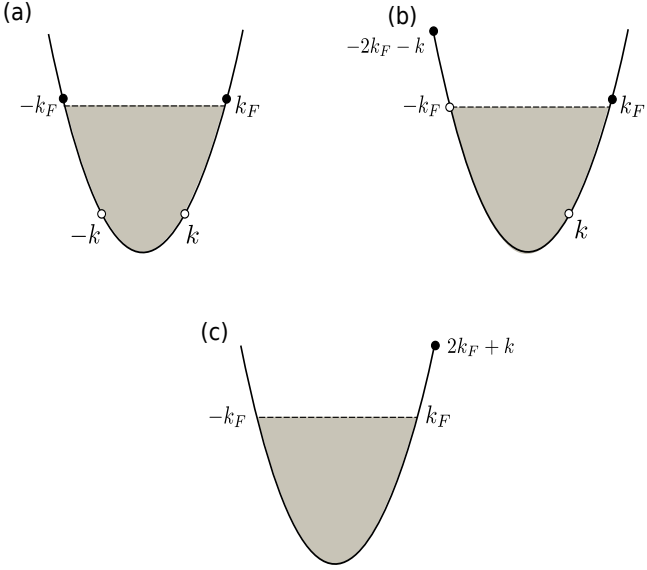}
\caption{\label{Fig2} The states (a) - (c) shown above are the only intermediate states that give non-zero contributions in the perturbative calculation of the form factors used to determine the prefactors $A_{0,+}(k), A_{1,-}(k)$ and $A_{1,+}(k)$, respectively.}
\end{figure}

For weakly interacting fermions we may directly evaluate form factors in the left hand sides of Eqs.~(\ref{fermion_scaling}),(\ref{density_scaling}), and (\ref{Ascaling}) using conventional perturbation theory and extract prefactors.
For example, since $\mu_{0,+}=-1+O(\hat{V}^2)$~\cite{Pustilnik2006Fermions, PRL_09}, one can expand the right hand side of Eq.~(\ref{Ascaling}) in powers of $\hat{V}$ as
\bea
\frac{(2\pi)^{3}A_{0,+}(k)}{L^{3}}\times[1+(\mu_{0,+}+1)\log(L/2\pi)+...]. \label{log_series}
\eea
While in an infinite size system this expansion is not convergent, for finite $L$ it is well defined
if one keeps $L$ finite but large, and then takes the limit $V(r)\to 0.$ We treat the interaction term perturbatively, and can e.g. write the expansion of the ground state as
\bea
|N\rangle = |{\rm FS}\rangle + \sum_{|\alpha\rangle}\frac{\langle \alpha | \hat V|{\rm FS}\rangle}{E_{{\rm FS}}-E_\alpha}|\alpha\rangle+...,
\eea
where $|{\rm FS}\rangle$ denotes a filled Fermi sea. Similar perturbative expressions can be written for the states on the left hand sides of Eqs.~(\ref{fermion_scaling}),(\ref{density_scaling}), and (\ref{Ascaling}), and one can then straightforwardly evaluate
the scaling of form factors with integer powers of $L$. Due to momentum constraints, only few intermediate states contribute
within lowest order perturbation theory. Eg. for $A_{0,+}(k)$ (see Fig.~\ref{Fig_def} a), the only sequence of states which contributes is the following:
first, $\hat V$ creates two particles at the right and left Fermi points, and two holes at $+k$ and $-k;$ (see Fig.~\ref{Fig2}a), second the operator $\hat \psi^{\dagger}(0)=\frac{1}{\sqrt{L}}\sum_{p}\hat \psi^{\dagger}(p)$ fills
in a hole at $+k,$ and we end up with the final state $|k,N+1\rangle$.

\subsection{Calculation of $A_{0,+}(k)$}
We expect the spectral function in the region $\omega >0, k \in (-k_F,k_F)$(see Fig.~\ref{Fig_def}a)  to behave as
\bea
& &A(k,\omega)\approx  2\pi A_{0,+}(k)\theta\left(\omega -\left( \frac{k_F^2 - k^2}{2m}+ \Delta\varepsilon\right)\right)\nonumber\\
& &\times\frac{(\omega -\left( \frac{k_F^2 - k^2}{2m}+ \Delta\varepsilon\right))^{-\mu_{0,+}}}{\Gamma(1-\mu_{0,+})(v+v_d)^{\mu_{0,+,L}}(v-v_d)^{\mu_{0,+,R}}},\label{A_0_p_def}\\
& & \Delta\varepsilon = \int_{-k_F}^{k_F}\frac{dk'}{2\pi}\left(V(k' - k) - V(k'-k_F)\right) + O(\hat{V}^2),\nonumber\eea
where $\Delta \varepsilon$ captures the shift in the dispersion curve of a non interacting fermionic system when interactions are introduced. To first order the shift is obtained by considering the Hartree-Fock corrections to the energy of eigenstates. 
The exponents $\mu_{0,+}, \mu_{0,+,L}, \mu_{0,+,R}$  (see Eq.~(\ref{mueqs}) and Ref.~\cite{muRLdef}) to leading order can be obtained using:
\bw
\bea
 \mu_{0,+,L} &=& \left(\frac{1}{\sqrt{K}} - \frac{\delta_-}{2\pi}\right)^2 = \left(1 + \frac{m(V(0)-V(2k_F))}{4\pi k_F} - \frac{m(V(k_F + k) - V(0))}{2\pi (k + k_F)} + O(\hat{V}^2)\right)^2,\nonumber\\
 \mu_{0,+,R} &=& \left(\frac{1}{\sqrt{K}} + \frac{\delta_+}{2\pi}\right)^2 = \left(1 + \frac{m(V(0)-V(2k_F))}{4\pi k_F} - \frac{m(V(k_F - k) - V(0))}{2\pi (k - k_F)}+  O(\hat{V}^2) \right)^2,\nonumber\\
\mu_{0,+} &=& 1 - \mu_{0,+,L} - \mu_{0,+,R}\nonumber\\
&=& -1 -\left(\frac{m(V(0)-V(2k_F))}{\pi k_F} +\frac{m(V(k-k_F)- V(0))}{\pi(k-k_F)} - \frac{m(V(k+k_F) - V(0))}{\pi (k+k_F)}\right) + O(\hat{V}^2). \\
\mu_{0,+} &=& -1 + O(\hat{V}).\label{mu0p}
\eea
\ew

Using \ceq{Ascaling} and the value of $\mu_{0,+}$ in the limit of vanishing interactions from \ceq{mu0p}, we may obtain the prefactor $A_{0,+}(k)$ from:
\bea
\label{a0pft}
A_{0,+}(k) &=& \frac{L^3}{(8\pi^3)}|\langle k,N+1|\hat{\psi}^\dagger(0)|N\rangle|^2.
\eea
Thus we need to evaluate the form factor
\be
\label{ffa0p}
\langle k, N+1|\hat{\psi}\dg(0)|N\rangle = \frac{1}{\sqrt{L}}\langle k, N+1|\sum_{p''} \pd_{p''}|N\rangle,
\ee
where $|k, N+1\rangle$ is the lowest energy eigenstate of the interacting system with $N+1$ particles and total momentum $k$ and $|N\rangle$ is the $N$ particle ground state. Since there is no exact answer for such a form factor for a generic interacting Fermi gas, we expand the ket vectors in Eq.~(\ref{ffa0p}) in terms of the eigenstates of the non interacting system perturbatively in $\hat{V}$ in the following way:
\bea
\label{ket_expansion}
|N\rangle &=& |{\rm FS}\rangle + \sum_{|\alpha\rangle}\frac{\langle \alpha|\hat{V}|{\rm FS}\rangle}{E_{{\rm FS}} - E_{\alpha}}|\alpha\rangle + O(\hat{V}^2),\nonumber\\
|k, N+1\rangle &=& |k, N+1\rangle^{(0)} + \sum_{|\alpha\rangle}\frac{\langle \alpha|\hat{V}|k, N+1\rangle^{(0)}}{E_{k,N+1} - E_{\alpha}}|\alpha\rangle \nonumber\\
&+& O(\V^2),
\eea
where, the state $|{\rm FS}\rangle$ is a filled Fermi sea of $N$ particles, and the state $|k, N+1\rangle^{(0)}$ is an eigenstate of the $N+1$ non interacting fermions, that has two particles with momenta $-k_F - 2\pi/L$ and $k_F+2\pi/L$, and a hole of momentum $-k$, on top of $|{\rm FS} \rangle$. From now on we will specify various states of the non-interacting system by describing the configuration of the state with respect to $|{\rm FS}\rangle$. The sum over $|\alpha\rangle$ runs over the entire Hilbert space of the non interacting system except the zeroth order state on the right, and $E_{\alpha}$ is the energy of the eigenstate $|\alpha\rangle$ of the non interacting system.

When we use the expansions in  \ceq{ket_expansion} to linear order in expression (\ref{ffa0p}) for the form factor, we obtain no contribution from terms that are zeroth order in $\V$ since it is not possible to create the state $|k,N+1\rangle^{(0)}$ by acting on $|{\rm FS}\rangle$ with just a single creation operator. When we consider the two terms generated by taking the zeroth order term from one of the ket vector expansions and the first order term from the other it becomes possible to obtain non zero contributions. Thus we are left considering two terms:
\bea
\label{cont1}
& &\sum_{|\alpha\rangle}\frac{1}{2L^{3/2}(E_{\rm FS} - E_{\alpha})} \langle k,N+1|^{(0)}\sum_{p''} \pd_{p''}|\alpha\rangle\nonumber\\
& &\times\langle \alpha |\sum_{q,p,p'} V(q)\pd_{p+q}\pd_{p'-q}\p_{p'}\p_{p}|{\rm FS} \rangle\\ 
&{\rm and}&\nonumber\\
\label{cont2}
& &\sum_{|\alpha\rangle}\frac{1}{2L^{3/2}(E_{k,N+1} - E_{\alpha})} \langle \alpha| \sum_{p''}\pd_{p''} |{\rm FS} \rangle\nonumber\\
& &\times \langle k,N+1|^{(0)}\sum_{q,p,p'} V(q)\pd_{p+q}\pd_{p'-q}\p_{p'}\p_{p}| \alpha\rangle.
\eea

We see that only the first term leaves a nonzero contribution for the following reasons. In term (\ref{cont1}), the matrix element of the momentum conserving interaction $\hat{V}$ limits the possibilities for $|\alpha\rangle$ to be only states with zero momentum. On the other hand, for the matrix element of the creation operator to be non zero we require that the state $|\alpha\rangle$ must have the same configuration as  $|k,N+1\rangle^{(0)}$ but with one more hole. Since the state $|k,N+1\rangle^{(0)}$ already has a hole, the only admissible state is one with two particles of momenta $k_F+2\pi/L$ and $-k_F+2\pi/L$ and two holes of momenta $k$ and $-k$ and is shown in Fig.~\ref{Fig2}(a). When we consider term (\ref{cont2}) we see that the states $|\alpha\rangle$ which will give non zero contributions must have total momenta $k$ and must contain one additional particle over the ground state. There can exist no such state since $|k| < k_F$, and consequently we may disregard term (\ref{cont2}).

Thus the only contributions we are left with are due to the term in expression (\ref{cont1}) for the state in Fig.~\ref{Fig2}(a):
\begin{widetext}
\bea
E_{\rm FS} - E_{2a} &=& \frac{(k_F+2\pi/L)^2}{M} - \frac{k^2}{M},\nonumber \\
\label{m3term}
\langle k,N+1|\hat{\psi}\dg(0)|N\rangle &=& \frac{M(V(k_F - k + 2\pi/L) -V(k_F+k+2\pi/L) - V(-k_F-k-2\pi/L) +V(-k_F+k-2\pi/L)) }{2L^{3/2}[(k_F+2\pi/L)^2-k^2]}\nonumber\\
 & &+ O(\V^2).
\eea
\end{widetext}

 Using the correspondence between the prefactor $A_{0,+}(k)$ and the form factor given in \ceq{a0pft} we obtain
\bea
\label{finite_size_scaling}
A_{0,+}(k) &=& \frac{L^3}{(8\pi^3)}|\langle k,N+1|\hat{\psi}\dg(0)|N\rangle|^2\nonumber\\
&=& \frac{M^2(V(k_F+k)-V(k_F-k))^2}{8\pi^3(k_F^2-k^2)^2},
\eea
to leading order in $\hat{V}$ and where we have used the fact that $V(r)$ is real and symmetric, thus $V(-q) = V(q)$. The divergence as $k \to \pm k_F$ from the denominator of \ceq{finite_size_scaling} still leads to a finite integral over $\omega>0$ when we substitute the expression for $A_{0,+}$ in \ceq{A_0_p_def}.
\bea\nonumber\eea
\bea\nonumber\eea
\bea\nonumber\eea

\subsection{Calculation of $A_{1,-}(k)$}

In the region $\omega <0, k \in (k_F,3k_F)$ (see Fig.\ref{Fig_def}a) the spectral function behaves as
\bea
\label{a1mscaling}
A(k,\omega)&\approx&  2\pi A_{1,-}(k)\nonumber\\
& &\times\left(\omega +\left( \frac{k_F^2 - (k-2k_F)^2}{2m} + \Delta \varepsilon\right)\right)^{-\mu_{1,-}}\nonumber\\
& &\times\frac{\theta\left(\omega + \left(\frac{k_F^2 - (k-2k_F)^2}{2m} + \Delta \varepsilon\right) \right)}{\Gamma(1-\mu_{1,-})(v+v_d)^{\mu_{1,-,L}}(v-v_d)^{\mu_{1,-,R}}},\nonumber\\
\Delta \varepsilon &=&  \int_{-k_F}^{k_F}\frac{dk'}{2\pi}\left(V(k' - k + 2k_F) - V(k'-k_F)\right)\nonumber\\
& & + O(\hat{V}^2).\label{A_1_m_def}
\eea
The exponents $\mu_{1,-}, \mu_{1,-,L}, \mu_{1,-,R}$ are given by \ceq{mueqs} and Ref.~\cite{muRLdef}, and can be written to leading order as
\bw
\bea
 \mu_{1,-,L} &=& \left(\sqrt{K} - \frac{\delta_-}{2\pi}\right)^2 = \left(1 - \frac{m(V(0)-V(2k_F))}{4\pi k_F} + \frac{m(V(k_F + k) - V(0))}{2\pi (k + k_F)} + O(\hat{V}^2)\right)^2,\nonumber\\
 \mu_{1,-,R} &=& \left(\sqrt{K} - \frac{\delta_+}{2\pi}\right)^2 = \left(1 - \frac{m(V(0)-V(2k_F))}{4\pi k_F} + \frac{m(V(k_F - k) - V(0))}{2\pi (k - k_F)}+  O(\hat{V}^2) \right)^2,\nonumber\\
\mu_{1,-} &=& 1 - \mu_{1,-,L} - \mu_{1,-,R}\nonumber\\
&=& -1 +\left(\frac{m(V(0)-V(2k_F))}{\pi k_F} -\frac{m(V(k-k_F)- V(0))}{\pi(k-k_F)} - \frac{m(V(k+k_F) - V(0))}{\pi (k+k_F)}\right) + O(\hat{V}^2). \\
\mu_{1,-} &=& -1 + O(\hat{V}).\label{mu1m}
\eea
\ew

 Using the value of the exponent in the absence of interactions and the finite size version of \ceq{a1mscaling} we may write
 \bea
A_{1,-}(k) =  \frac{L^3}{(8\pi^3)}|\langle k,N-1|\hat{\psi}(0)|N\rangle|^2.
 \eea
Consequently we need the matrix element
\bea
\label{ffa1m}
\langle k,N-1|\p(0)|N\rangle &=& \langle 2k_F+k_h,N-1|\p(0)|N\rangle\nonumber\\
 &=& \frac{1}{\sqrt{L}}\langle 2k_F+k_h|\sum_{p''} \p_{p''}|N\rangle, \nonumber\\
\eea
where $k_h = k - 2k_F \in (-k_F,k_F),$ and the state $|2k_F+k_h,N-1\rangle$ is the lowest energy eigenstate of momentum $2k_F+k_h$ of $N-1$ interacting fermions and $|N\rangle$ is the $N$ particle ground state of the system. Again we perform a perturbative expansion of the ket vectors in orders of $V(q)$ in eigenstates of the free fermions as in \ceq{ket_expansion}. We note that the unperturbed version of the excited state,  $|2k_F+k_h, N-1\rangle^{(0)}$, will contain two holes with momenta $k_F, k_h$ and a particle of momentum $-k_F$.

We find that there is no zeroth order contribution since the state $|2k_F+k_h, N-1\rangle^{(0)}$ cannot be obtained by the action of a single annihilation operator on the ground state, since it contains two holes and a particle over the ground state. Thus we look to the first order terms of the form given in expressions (\ref{cont1}) and (\ref{cont2}). In order for the linear order terms to give a non-zero contribution we need the matrix elements of $\hat{V}$, which can only connect states of the same momentum, and that of the annihilation operator between $\langle \alpha|$ and $|{\rm FS} \rangle$, to simultaneously be non-zero. The first requirement automatically narrows down our choices to two subsets of intermediate states $\{|\alpha\rangle\}$. One subset contains states of zero momentum while the other contains states of momentum $2k_F +k_h$. For the matrix element of the annihilation operator to be non-zero, the states $|\alpha\rangle$ of momentum $2k_F + k_h$ must be identical to the filled Fermi sea but with a single additional hole - however this is impossible since the momentum of such a state can at most be $k_F$. Thus we must focus our attention on the term
\bea
& &\sum_{|\alpha\rangle}\frac{1}{2L^{3/2}(E_{\rm FS} - E_{\alpha})} \langle 2k_F +k_h, N-1 |\sum_{p''} \p_{p''}|\alpha\rangle\nonumber\\
&\times&\langle \alpha |\sum_{q,p,p'} V(q)\pd_{p+q}\pd_{p'-q}\p_{p'}\p_{p}|{\rm FS} \rangle. \nonumber
\eea
The only state $|\alpha\rangle$ that gives a contribution to this term contains two holes of momenta $k_F$ and $k_h$ and two particles of momenta $-k_F-2\pi/L$ and $2k_F + k_h+2\pi/L$, and is shown in Fig.~\ref{Fig2}(b). Consequently the matrix element to first order in $\V$ can be calculated as follows:
\begin{widetext}
\bea
E_{\rm FS} - E_{2b} &=& -\frac{(2k_F+k_h+2\pi/L)^2}{2M} - \frac{(k_F+2\pi/L)^2}{2M} + \frac{k_F^2}{2M} + \frac{k_h^2}{2M} = -\frac{2k_F}{M}\left(k_F + k_h \right) + O(1/L).\nonumber\\
\langle 2k_F+k_h,N-1|\hat{\psi}(0)| N \rangle &=& \frac{M(V(2k_F) - V(k_F+k_h))}{2k_FL^{3/2}(k_F+k_h)} + O(\V^2,1/L).
\eea
\end{widetext}

We may then express the prefactor $A_{1,-}(k)$ in terms of the matrix element using a similar correspondence to \ceq{finite_size_scaling}:
\bea
\label{a1mdivergence}
A_{1,-}(k) &=& \frac{L^3}{(8\pi^3)}|\langle k,N-1|\hat{\psi}(0)|N\rangle|^2 \nonumber\\
&=& \frac{M^2(V(2k_F)-V(k_F-k))^2}{32\pi^3k_F^2(k_F-k)^2},
\eea
where we have substituted $k_h = k - 2k_F$ to obtain the final answer in terms of $k_F \leq k \leq 3k_F$. 

As in the case of $A_{0,+}(k)$, we note again that the divergence as $k \to k_F$ from the denominator of \ceq{a1mdivergence} still leads to a finite value when we substitute the expression for $A_{1,-}(k)$ in \ceq{A_1_m_def} and integrate over $\omega < 0$.

\subsection{Calculation of $A_{1,+}(k)$}
When  $\omega >0, k \in (k_F,3k_F)$ (see Fig.~\ref{Fig_def}a) the spectral function behaves as
\bea
\label{a1pscaling}
A(k,\omega)&\approx&  2\pi A_{1,+}(k)\nonumber\\
& &\left(\omega - \left(\frac{k_F^2 - (k-2k_F)^2}{2m} + \Delta\varepsilon\right)  \right)^{-\mu_{1,+}}\nonumber\\
& &\times\frac{\theta\left(\omega - \left(\frac{k_F^2 - (k-2k_F)^2}{2m} + \Delta\varepsilon\right)  \right)}{\Gamma(1-\mu_{1,+})(v+v_d)^{\mu_{1,+,L}}(v-v_d)^{\mu_{1,+,R}}},\nonumber\\
\Delta \varepsilon&=&  \int_{-k_F}^{k_F}\frac{dk'}{2\pi}\left(V(k' - k + 2k_F) - V(k'-k_F)\right)\nonumber\\
& & + O(\hat{V}^2).
\eea
Moreover we obtain perturbative expressions for the exponents $\mu_{1,+,R}, \mu_{1,+,L}$ and $\mu_{1,+}$ from \ceq{mueqs} and Ref.~\cite{muRLdef}:
\bw
\bea
 \mu_{1,+,L} &=& \left(\sqrt{K} + \frac{1}{\sqrt{K}} - \frac{\delta_-}{2\pi}\right)^2 = \left(2 - \frac{m(V(k_F + k) - V(0))}{2\pi (k + k_F)} + O(\hat{V}^2)\right)^2,\nonumber\\
 \mu_{1,+,L} &=& \left(\sqrt{K} - \frac{1}{\sqrt{K}} - \frac{\delta_+}{2\pi}\right)^2 = \left(\frac{m(V(0)-V(2k_F))}{4\pi k_F} + \frac{m(V(k_F - k) - V(0))}{2\pi (k - k_F)}+  O(\hat{V}^2) \right)^2,\nonumber\\
\mu_{1,+} &=& 1 - \mu_{1,-,L} - \mu_{1,-,R}\nonumber\\
&=& -3 +\frac{2m(V(k+k_F) - V(0))}{\pi (k+k_F)} + O(\hat{V}^2). \\
\mu_{1,+} &=& -3 + O(\hat{V}).\label{mu1p}
\eea
\ew

Using the value of the exponent in the absence of interactions and the finite size version of \ceq{a1pscaling} we may write
 \bea
A_{1,+}(k) =  \frac{L^5}{(32\pi^5)}|\langle k,N-1|\hat{\psi}^{\dagger}(0)|N\rangle|^2.
 \eea
Thus we first focus on the matrix element
\bea
\label{ffa1p}
\langle k,N+1|\hat{\psi}\dg(0)|N\rangle&=&\langle 2k_F - k_h,N+1|\hat{\psi}\dg(0)|N\rangle\nonumber\\
&=& \frac{1}{\sqrt{L}}\langle 2k_F - k_h|\sum_{p''} \pd_{p''}|N\rangle,\nonumber\\
\eea
where $k_h = 2k_F - k \in (-k_F, k_F)$, and the state $|2k_F - k_h,N+1\rangle$ is the lowest energy eigenstate of the $N+1$ particle interacting system with momentum $2k_F - k_h$ and $|N\rangle$ is the $N$ particle ground state. We again expand the ket vectors in \ceq{ffa1p} as in \ceq{ket_expansion}. The corresponding eigenstate of the free fermions is $|2k_F - k_h, N+1\rangle^{(0)}$, an $N+1$ particle state with two particles with momenta $k_F + 2\pi/L$ and  $k_F + 4\pi/L$, and one hole with momentum $k_h$. Here too we must perturbatively expand in the states for which we calculate the above matrix element. There is no zeroth order contribution since the state $|2k_F-k_h,N+1\rangle^{(0)}$ cannot be created by the action of a single creation operator on the ground state.

Upon considering the first order terms of the form given in expressions (\ref{cont1}), (\ref{cont2}) we find the only non-zero contribution is to the term
\bea
& &\sum_{|\alpha\rangle}\frac{1}{2L^{3/2}(E_{2k_F-k_h,N+1} - E_{\alpha})} \langle \alpha |\sum_{p''} \pd_{p''}|{\rm FS}\rangle\nonumber\\
&\times&\langle 2k_F-k_h,N+1|\sum_{q,p,p'}V(q)\pd_{p+q}\pd_{p'-q}\p_{p'}\p_{p}|\alpha\rangle, \nonumber
\eea
from the state depicted in Fig.~\ref{Fig2}(c). Due to the kinematic constraint imposed by $\hat{V}$, the states that may contribute to the matrix element must either have zero momentum and must be connected to the state $|2k_F-k_h,N+1\rangle^{(0)}$ by the action of the creation operator, or have momentum $2k_F - k_h$ and must be connected to the ground state by the action of the annihilation operator. The latter condition is the only kinematically feasible one, and the only allowed state is the one in Fig.~\ref{Fig2}(c).

Thus we may collect all the terms that are linear order in $\V$ as follows:
\begin{widetext}
\bea
E_{2k_F-k_h,N+1} - E_{2c} &=&  \frac{(k_F+4\pi/L)^2}{2M}+\frac{(k_F+2\pi/L)^2}{2M} - \frac{k_h^2}{2M} - \frac{(2k_F-k_h+6\pi/L)^2}{2M} = -\frac{(k_F -k_h)^2}{M} + O(1/L),\nonumber \\
\langle 2k_F-k_h,N+1|\hat{\psi}\dg(0)|N\rangle &=& \frac{M(V(k_F - k_h + 4\pi/L) - V(k_F-k_h +2\pi/L))}{L^{3/2}(k_F-k_h)^2}  = \frac{2\pi MV'(k_F-k_h)}{L^{5/2}(k_F - k_h)^2}.
\eea
\end{widetext}
Note that in the above expression, the appearance of the derivative is essential to give the correct power of $1/L$, which is evident when one considers the correspondence between the prefactor $A_{1,+}(k)$ and the form factor (\ref{ffa1p}) given below.

From finite size scaling the prefactor $A_{1,+}(k)$ can be obtained using the matrix element above:
\be
\label{a1pdiv}
A_{1,+}(k) = \frac{L^5}{32\pi^5}|\langle k,N+1|\hat{\psi}\dg(0)|N\rangle|^2 = \frac{M^2 V'(k-k_F)^2}{8\pi^3 (k - k_F)^4},
\ee
where we have substituted $k_h = 2k_F - k$ to obtain the final answer in terms of $k_F \leq k \leq 3k_F.$ We note that the divergence of \ceq{a1pdiv} as $k \to k_F$ is generically the same $\propto (k-k_F)^{-2}$ divergence seen in $A_{1,-}(k\to k_F)$. Yet again the integral of the spectral weight over $\omega > 0$ remains finite as can be seen from \ceq{a1pscaling}.

 The procedure to obtain the prefactors $A_{n \geq 1, \pm}(k)$ is similar to the one used so far. In general, to lowest non-vanishing order $A_{n\geq1,\pm}(k)$ will be $\propto V^{2n}$,
but one needs to sum contributions to the form factor expansion from a rapidly growing number of intermediate states. Moreover terms appearing with higher orders of $\hat{V}$ need to be carefully separated from terms $\propto$ (log$(L))^n$ that generate corrections to the exponent (see Eq.~(\ref{log_series})). The latter procedure is demonstrated in Sec.~IIIF where we calculated $A_{0,-}(k)$ to the lowest non-vanishing order in $\hat{V}$.
 \subsection{Calculation of prefactor $C_1$}
  \begin{figure}
\includegraphics[width=8.5cm,height=4cm]{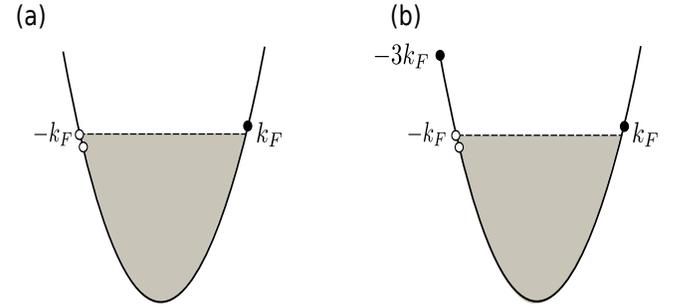}
\caption{\label{C1states} The state in (a) depicts the parent state with respect to which we calculate the annihilation operator form factor to obtain prefactor $C_1$. That state in (b) depicts the only intermediate state which gives non-zero contribution to the form factor in the perturbative expansion to first order in $\V$.}
\end{figure}

 We now consider the prefactor $C_1$ defined in Eq.~(\ref{Cmdef}). We denote by $|m=1,N-1\rangle$ the lowest energy $N-1$ particle eigenstate of the interacting fermion system of momentum $3k_F$. The corresponding state of the free fermion system (zeroth order term of the perturbative expansion) has two holes at the left Fermi point and one particle at the right Fermi point and is depicted in Fig.~\ref{C1states}(a). We wish to calculate the annihilation operator form factor for this state and the $N$ particle ground state $|N\rangle$:
 \be
 \langle m = 1,N-1|\hat{\psi}(0)|N\rangle = \frac{1}{\sqrt{L}}\langle m=1,N-1|\sum_{p''}\p_{p''}|N\rangle.
 \ee

 We may again expand the states above like \ceq{ket_expansion}. We find that the first non-zero contributions in the perturbative expansion appear at the first order in $\V$. To this order we obtain the following two terms by expanding the states in the bra- and the ket vector perturbatively in $\V$:
 \bea
 \label{first term}
& &\sum_{|\alpha\rangle}\frac{1}{2L^{3/2}(E_{\rm FS} - E_{\alpha})} \langle m=1,N-1|\sum_{p''}\p_{p''}|\alpha\rangle\nonumber\\
& &\langle \alpha |\sum_{q,p,p'} V(q)\p\dg_{p+q}\p\dg_{p'-q}\p_{p'}\p_{p}|{\rm FS} \rangle,\\
& &{\rm and}\nonumber\\
\label{second term}
& &\sum_{|\alpha\rangle}\frac{1}{2L^{3/2}(E_{m=1,N-1} - E_{\alpha})} \langle \alpha| \sum_{p''}\p_{p''}|{\rm FS} \rangle \nonumber\\
& &\langle m=1,N-1|\sum_{q,p,p'} V(-q)\p\dg_{p+q}\p\dg_{p'-q}\p_{p'}\p_{p}| \alpha\rangle.\nonumber\\
\eea

 The states which give non-zero contributions to the term (\ref{first term}) must have 0 momentum. Furthermore we must be able to transform them into the state $|m=1,N-1\rangle$ by the action of a single annihilation operator. This leaves only one possibility shown in Fig.~\ref{C1states}(b). While for the term (\ref{second term}) to be non-zero we require states which have momentum $3k_F$ which can be obtained from the ground state by the action of a single annihilation operator, i.e. a state with one hole of momentum $-3k_F$, which is impossible to achieve.

Thus we find the form factor as follows:
\begin{widetext}
\bea
E_{\rm FS}-E_{3b} &=& \frac{k_F^2}{2M} + \frac{(k_F -2\pi/L)^2}{2M} - \frac{(k_F + 2\pi/L)^2}{2M} - \frac{(3k_F)^2}{2M} = -\frac{4k_F}{M}\left(k_F + \frac{\pi}{L}\right),\nonumber\\
 \langle m = 1,N-1|\hat{\psi}(0)|{\rm FS}\rangle  &=& \frac{M(V(2k_F + 2\pi/L) - V(2k_F))}{4L^{3/2}k_F^2} = \frac{M\pi V'(2k_F)}{2L^{5/2}k_F^2}.
\eea
\end{widetext}

We may express $C_1$ in terms of this form factor using the correspondence we have derived between the form factor calculated above and our desired prefactor:
\be
C_1 = \frac{2(\rho_0 L)^5}{(2\pi)^5} |\langle m = 1,N-1|\hat{\psi}(0)|N\rangle|^2 = \frac{M^2\rho_0 V'(2k_F)^2}{64\pi^7}.
\ee

\subsection{Calculation of prefactor $A_2$}
To obtain the prefactor $A_2$ in Eq.~(\ref{Amdef}) we must calculate the following matrix element:
\be
\label{matelem}
\langle m=2, N|\hat{\rho}(0)|N\rangle=\langle m=2, N|\frac{1}{L}\sum_{q',k'} \pd_{k'+q'} \p_{k'}|N\rangle,
\ee
where $|m=2, N\rangle$ is the lowest energy eigenstate of the interacting system with momentum $4k_F$. The unperturbed version of this state, for $N$ free fermions, contains two adjacent holes in the vicinity of the left Fermi point and two adjacent particles in the vicinity of the right Fermi point.

We evaluate the form factor in \ceq{matelem} by again perturbatively expanding the ket vectors in the free fermion basis, in powers of $\V$ as in \ceq{ket_expansion}. We again disregard the terms that are zeroth order in $\hat{V}$ since the density operator cannot create the $m=2$ Umklapp state from the ground state, and consider the first order terms for the first non vanishing contributions.

\begin{figure}
\includegraphics[width=8.5 cm,height=7.5cm]{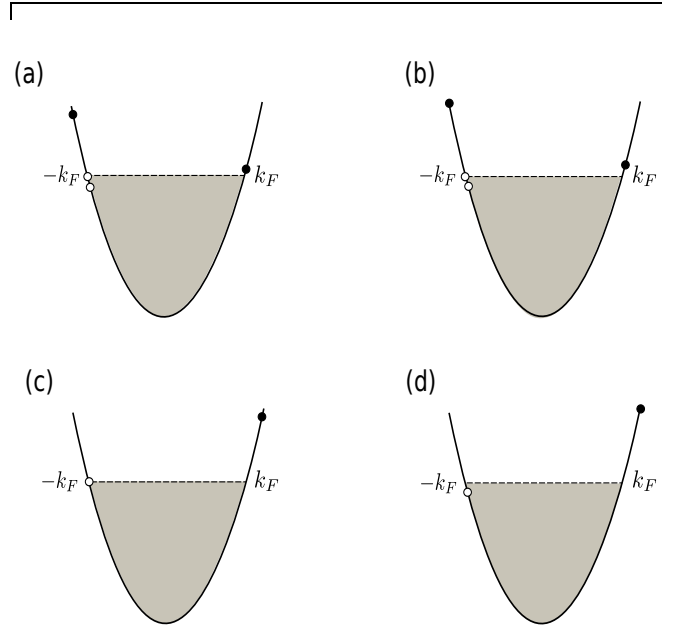}
\caption{\label{Fig_A2} The states (a)-(d) indicated above are the only intermediate states that give non-zero contribution to the first order in $\hat{V}$ in the perturbative expansion of the form factor in Eq.~(\ref{matelem}). This form factor is used to determine $A_2$ to leading oder in $\hat{V}$. Contributions to the form factor from the states (a) and (b) pictured above are calculated in Eq.~(\ref{m4ab}) and the contributions from states (c) and (d) in Eq.~(\ref{m4cd}).}
\end{figure}

The first term of $O(\hat{V}$) is 
\bea
\label{t1}
& &\sum_{|\alpha\rangle}\frac{1}{2L^2(E_{\rm FS} - E_{\alpha})} \langle m=2,N|\sum_{q',k'}\pd_{k'+q'} \p_{k'}|\alpha\rangle\nonumber\\
& &\langle \alpha |\sum_{q,p,p'} V(q)\pd_{p+q}\pd_{p'-q}\p_{p'}\p_{p}|{\rm FS} \rangle.
\eea

The constraint from the first matrix element tells us that the intermediate state can differ from the Umklapp state $|m=2,N\rangle$ in only two ways since the density operator can either create a new particle-hole pair, or move an existing hole or particle to a different `spot' in momentum space.

 At this stage there are two possible types of intermediate states $|\alpha\rangle$. One possibility is a state with only one particle-hole pair on top of the ground state such that this pair is one of the ones contained in the configuration $|m=2, N\rangle$. In this case the density operator creates the second excitation of roughly $p \approx 2k_F$. The second possible type is an intermediate state which contains two particle-hole pairs. Such a state can at most differ by one particle or hole from the configuration $|m=2, N\rangle$ since we can use the density operator to move this particle or hole to give the correct final configuration.

In the case where $|\alpha\rangle$ contains only a single particle-hole pair, the only way the second matrix element can be non zero is if  the action of the $\hat{V}$ creates one particle hole pair of momentum $p \approx 2k_F$ and the other of momentum 0. This cannot be done since the operator is explicitly momentum conserving. Thus we must only consider intermediate states with two particle hole pairs where one particle or hole may differ from the configuration of $|m=2, N \rangle$.

 We can further narrow down the intermediate state $|\alpha\rangle$ to the two shown in Fig.~\ref{Fig_A2}(a), (b). The reasoning is the following; let us try to generate a state that differs by one particle from the desired Umklapp state. To do so we use one of the creation-annihilation pairs to generate a hole at either $-k_F$ or $-k_F +2\pi/L$ and a particle at either $k_F+2\pi/L$ or $k_F +4\pi/L$. The second annihilation operator must then make the hole that was not made by the first. This choice along with the momentum conserving constraint fixes the last particle to be at approximately $-3k_F$. The above procedure will generate eight terms - for each of the states in Fig.~\ref{Fig_A2}(a), (b), there are two choices for where one of the creation operators can act and two independent choices for where one of the annihilation operators can act. The second creation and annihilation operators are fixed. This corresponds to a total of four ways to connect the ground state to each of the two states, giving a total of eight terms.

  If on the other hand we try to generate a state that differs by one hole from the Umklapp state, we find that there is no way to do so given the kinematic constraints imposed by the interaction term, i.e. using a similar argument to the case of the deviant particle, we find for the state differing by one hole the only way to conserve momentum is to have the annihilation operator act outside the Fermi sea. Thus we exhaust all the possible intermediate states that give  non zero contributions for both the matrix elements.

  We may collect the various contributions to \ceq{t1} for the states shown in Fig.~\ref{Fig_A2} (a) and (b):
\begin{widetext}
 \bea
 E_{\rm FS} - E_{4a} &=& \frac{-k_F^2}{2M} + \frac{(-k_F+2\pi/L)^2}{2M} - \frac{(k_F+2\pi/L)^2}{2M} - \frac{(-3k_F)^2}{2M} = -\frac{4k_F}{M}\left(k_F + \frac{\pi}{L}\right),\nonumber\\
 M_{4a} &=& -\frac{M (V(2k_F+\frac{2\pi}{L})- V(2k_F))}{8L^2k_F(k_F+\frac{\pi}{L})} -\frac{M(V(-2k_F-\frac{2\pi}{L}) - V(-2k_F))}{8L^2k_F(k_F+\frac{\pi}{L})},\nonumber\\
  E_{\rm FS}-E{4b} &=& \frac{(-k_F)^2}{2M} + \frac{(-k_F+2\pi/L)^2}{2M} - \frac{(k_F+4\pi/L)^2}{2M} - \frac{(-3k_F-2\pi/L)^2}{2M} =  -\frac{4k_F}{M}\left(k_F + \frac{3\pi}{L}+\frac{2\pi}{k_F L^2}\right),\nonumber \\
  M_{4b} &=& -\frac{M( V(2k_F + \frac{2\pi}{L}) - V(2k_F + \frac{4\pi}{L})) }{8L^2k_F\left(k_F + \frac{3\pi}{L}+\frac{2\pi^2}{k_F L^2}\right)}-\frac{M (V(-2k_F - \frac{2\pi}{L})-V(-2k_F - \frac{4\pi}{L}))}{8L^2k_F\left(k_F + \frac{3\pi}{L}+\frac{2\pi^2}{k_F L^2}\right)}.\label{m4ab}
\eea
\end{widetext}
\bea
\nonumber
\eea
\bea
\nonumber
\eea
 Next we consider the second term that is linear order in $V(q):$
 \bea
 \label{t2}
 & &\sum_{|\alpha\rangle}\frac{1}{2L^2(E_{m=2,N} - E_{\alpha})} \langle \alpha| \sum_{q',k'} \pd_{k'+q'} \p_{k'}|{\rm FS} \rangle\nonumber\\
 & & \langle m=2, N|\sum_{q,p,p'} V(-q)\pd_{p+q}\pd_{p'-q}\p_{p'}\p_{p}| \alpha\rangle.
 \eea

 We find that for the first matrix element to be non zero,  $|\alpha\rangle$ must contain only one particle hole excitation over the ground state which the density operator will remove. Furthermore, because $\hat{V}$ conserves momentum, $|\alpha\rangle$ must have momentum $4k_F$. The interaction term must act on the state $|\alpha\rangle$ in the following way; one of the pairs of creation-annihilation operators must create one of the paired particle-hole excitations of momentum $\approx 2k_F$ as found in $|m=2, N\rangle$. The other is automatically constrained to change the momentum of a particle by $\approx 2k_F$. The combined action of the four fermion operators on $|\alpha\rangle$ must generate $|m=2,N\rangle$ in order for the matrix element to be non zero. The allowed intermediate states that give a non zero contribution from the term (\ref{t2}) are illustrated in Fig.~\ref{Fig_A2} (c), (d), and give the following contributions:
\begin{widetext}
 \bea
 E_{m=2,N} - E_{4c} &=& \frac{(k_F+2\pi/L)^2}{2M} + \frac{(k_F+4\pi/L)^2}{2M} - \frac{(-k_F)^2}{2M} - \frac{(-k_F + 2\pi/L)^2}{2M} + \frac{(-k_F)^2}{2M} - \frac{(3k_F+ 4\pi/L)^2}{2M} \nonumber\\
 &=& -\frac{4k_F}{M}\left(k_F+\frac{\pi}{L}\right),\nonumber \\
 M_{4c} &=&  -\frac{M (V(2k_F+\frac{2\pi}{L})- V(2k_F))}{8L^2k_F(k_F+\frac{\pi}{L})} -\frac{M (V(-2k_F-\frac{2\pi}{L}) - V(-2k_F))}{8L^2k_F(k_F+\frac{\pi}{L})},\nonumber \\
 E_{m=2,N} - E_{1d} &=& \frac{(k_F+2\pi/L)^2}{2M} + \frac{(k_F+4\pi/L)^2}{2M} - \frac{(-k_F)^2}{2M} - \frac{(-k_F + 2\pi/L)^2}{2M} + \frac{(-k_F+2\pi/L)^2}{2M} - \frac{(3k_F+ 6\pi/L)^2}{2M} \nonumber\\
 &=& -\frac{4k_F}{M}\left(k_F+\frac{3\pi}{L}+\frac{2\pi^2}{k_FL^2}\right),\nonumber \\
 M_{4d} &=&  -\frac{M (V(2k_F+\frac{2\pi}{L})- V(2k_F+\frac{4\pi}{L}))}{8L^2k_F(k_F+\frac{3\pi}{L}+\frac{2\pi^2}{k_FL^2})} -\frac{M (V(-2k_F-\frac{2\pi}{L})+ V(-2k_F-\frac{4\pi}{L}))}{8L^2k_F(k_F+\frac{3\pi}{L}+\frac{2\pi^2}{k_FL^2})}.\nonumber\\\label{m4cd}
 \eea
\end{widetext}
 Collecting these terms gives the following result to leading order in $1/L, \V$
 \be
\langle m=2, N|\hat{\rho}(0)|GS\rangle= -\frac{2M\pi^2}{k_F^3L^4}\left[V'(2k_F)-k_FV''(2k_F)\right].\nonumber
 \ee

 This means that for  $A_2$ we have
 \bea
 A_2 &=& \frac{2(\rho_0 L)^8}{(2\pi)^8\rho_0} |\langle m=2, N|\hat{\rho}(0)|GS\rangle|^2\nonumber\\
 &=& \frac{M^2\rho_0}{32\pi^{10}}\left[V'(2k_F)-k_FV''(2k_F)\right]^2.
 \eea

\subsection{Calculation of prefactor $A_{0,-}(k)$}

We can identify the prefactor $A_{0,-}(k)$ using:
\bw
\bea
\label{log_expansion}
|\langle k,N-1|\p(0)|N\rangle|^2 &=& \frac{2\pi A_{0,-}(k)}{L}\left(\frac{2\pi}{L}\right)^{1-\mu_{0,-}} = \frac{2\pi A_{0,-}}{L}\left(1 + (1-\mu_{0,-})\log\left(\frac{2\pi}{L}\right) + ... \right)\nonumber\\
&=&\frac{2\pi}{L}\left(1+ A_{0,-}^{(2)} + O(\V^{4})\right)\left(1+(1-\mu_{0,-}^{(2)})\log\left(\frac{2\pi}{L}\right) + O(\log^2, \V^4)\right)\nonumber\\
&=&\frac{2\pi}{L}\left(1+A_{0,-}^{(2)} +(1-\mu_{0,-}^{(2)})\log\left(\frac{2\pi}{L}\right)  + O(\log^2, \V^4)\right),\nonumber\\
\mu_{0,-}^{(2)} &=& 1- \frac{m^2}{4\pi^2}\frac{[V(0)-V(k_F+k)]^2}{(k_F+k)^2} - \frac{m^2}{4\pi^2}\frac{[V(0)-V(k_F-k)]^2}{(k_F-k)^2},
\eea
\ew
where in the second line of the above expression the first pair of parentheses contains the perturbative expansion of the prefactor and the second contains the expansion of the power law. The superscript (2) on a term indicates that the term is of $O(\V^2)$ accuracy.

Similarly to Eq.~(\ref{Ascaling}), the form factor required to obtain $A_{0,-}(k)$ is
\be
\label{ffa0m}
\langle k,N-1|\p(0)|N\rangle = \frac{1}{L^{1/2}}\langle k,N-1|\sum_{p''}\p_{p''}|N\rangle.
\ee
The state $|k,N-1\rangle$ contains a single hole of momentum $k$. The perturbative calculation of the form factor in \ceq{ffa0m} is complicated by the fact that at the zeroth order, i.e. for the non-interacting Fermi gas, the form factor is already non-zero. Consequently any dependence on the pair potential $\hat{V}$ enters as a term beyond leading order. This is problematic because terms that arise after the first non-vanishing order in the expansion of a generic parent form factor make two types of contributions. They not only contribute sub-leading corrections to the prefactor, which we require, but also generate powers of $\log(L)$ which give rise to the non-trivial power law in $L$.

Let us first expand the ket vectors in \ceq{ffa0m} up to second order in $\hat{V}$ and then show that this is required because there are no $O(\hat{V})$ contributions to the form factor.
\begin{widetext}
\bea
\label{second_order_expansion}
|k, N-1\rangle &=& |k, N-1\rangle^{(0)} + \sum_{\al}\frac{\langle \alpha|\V|k, N-1\rangle^{(0)}}{E_{k,N-1} - E_{\alpha}}\al + \sum_{\al,|\beta\rangle}\frac{\langle \alpha|\V|\beta\rangle\langle \beta|\V|k,N-1\rangle^{(0)}}{(E_{k,N-1}-E_{\alpha})(E_{k,N-1} - E_{\beta})}\al \nonumber\\
& & - \sum_{\al} \frac{\langle k,N-1|^{(0)}\V|k,N-1\rangle^{(0)}\langle\alpha|\V|k,N-1\rangle^{(0)}}{(E_{k,N-1} - E_{\alpha})^2}\al-|k,N-1\rangle^{(0)}\sum_{\al} \frac{|\langle \alpha|\V|k,N-1\rangle^{(0)}|^2}{2(E_{k,N-1} - E_\alpha)^2} ,\nonumber\\
|N\rangle &=&|{\rm FS} \rangle + \sum_{\al}\frac{\langle \alpha|\V|{\rm FS} \rangle}{E_{\rm FS} - E_{\alpha}}\al + \sum_{\al,|\beta\rangle}\frac{\langle \alpha|\V|\beta\rangle\langle \beta|\V|{\rm FS}\rangle}{(E_{\rm FS}-E_{\alpha})(E_{\rm FS} - E_{\beta})}\al - \sum_{\al} \frac{\langle {\rm FS}|\V|{\rm FS}\rangle\langle\alpha|\V|\rm FS \rangle}{(E_{k,N-1} - E_{\alpha})^2}\al\nonumber\\
& &-|{\rm FS}\rangle \sum_{\al}\frac{|\langle \alpha|\V|{\rm FS}\rangle|^2}{2(E_{\rm FS} - E_{\alpha})^2}.
\eea
\end{widetext}

Clearly there is a trivial zeroth order contribution,
\be
\frac{1}{L^{1/2}}\langle k,N-1|^{(0)}\p_{k}|{\rm FS}\rangle = \frac{1}{L^{1/2}},
\ee
where the annihilation operator removes the particle of momentum $k$ from the ground state to connect it to $|k,N-1\rangle^{(0)}$. 

Furthermore there can be no $O(\V)$ terms. There are two sources of first order terms corresponding to keeping a first order term in one of the ket vector expansions and the zeroth order term in the other. Moreover, due to the momentum conservation constraint imposed by $\V$ the intermediate states corresponding to these two groups of terms will have a total momentum of either 0 (same as $|{\rm FS}\rangle$) or $-k$ (same as $|k,N-1\rangle^{(0)}$). Finally, these intermediate states must be connected to $|k,N-1\rangle^{(0)}$ or $|{\rm FS}\rangle$ respectively, by the action of a single annihilation operator. The only zero momentum state that can be connected to $|k, N-1\rangle^{(0)}$ by a single annihilation operator is $|{\rm FS}\rangle$ since the annihilation operator has to remove momentum $|k| < k_F$ by creating a single hole, necessarily carrying momentum $k$. On the other hand the same argument holds for intermediate states of momentum $-k$ connecting the ground state by the action of $\p$ on the left, i.e. the only state that can be connected to $|{\rm FS}\rangle$ in this way is $|k,N-1\rangle^{(0)}$. Since the sum over intermediate states specifically forbids them from being identical to the unperturbed state, we obtain no contribution of $O(\hat{V})$ to the form factor. 

There are three ways of generating $O(\V^2)$ terms: we may either pick $O(\V)$ terms in the ket vector expansions of both states in \ceq{ffa0m}, or pick an $O(\V^2)$ term from one ket vector and an unperturbed term from the other (there are two ways of doing this). However not every term will give a contribution. Some terms can be discarded for the same reason the $O(\V)$ terms drop out. This is because the structure of the matrix elements appearing in these terms is the same as in the $O(\hat{V})$ terms - i.e. they require the matrix element of the annihilation operator between either $|{\rm FS}\rangle$ or $|k,N-1\rangle^{(0)}$ and an intermediate state to be zero with the constraint that the intermediate state must have the same momentum as, but cannot be equal to either of those states. Thus we need only consider the following terms:
\bw
\bea
\label{f1}
F_1 &=& \frac{\displaystyle \sum_{\al, |\beta\rangle} \sum_{p''}\sum_{p_1,p'_1,q_1}\sum_{p_2,p'_2,q_2}\langle \alpha|\p_{p''}|\beta\rangle\langle k,N-1|^{(0)}V(q_1)\pd_{p_1 +q_1}\pd_{p'-q_1}\p_{p'_1}\p_{p_1}|\alpha\rangle\langle\beta|V(q_2)\pd_{p_2+q_2}\pd_{p'_2-q_2}\p_{p'_2}\p_{p_2}|{\rm FS}\rangle}{4L^{5/2}(E_{k,N-1} - E_\alpha)(E_{\rm FS} - E_\beta)},\nonumber\\ \\
\label{f2}
F_2 &=& -\frac{\displaystyle \sum_{\al} \sum_{p''}\sum_{p_1,p'_1,q_1}\langle k,N-1|^{(0)}\p_{p''}|{\rm FS} \rangle| \langle {\rm FS} |V(q_1)\pd_{p_1 +q_1}\pd_{p'-q_1}\p_{p'_1}\p_{p_1}\al|^2}{8L^{5/2}(E_{\rm FS} - E_\alpha)^2},\\
\label{f3}
F_3 &=& -\frac{\displaystyle \sum_{\al} \sum_{p''}\sum_{p_1,p'_1,q_1}\langle k,N-1|^{(0)}\p_{p''}|{\rm FS}\rangle |\langle k,N-1 |^{(0)}V(q_1)\pd_{p_1 +q_1}\pd_{p'-q_1}\p_{p'_1}\p_{p_1}|\alpha\rangle|^2}{8L^{5/2}(E_{k,N-1} - E_\alpha)^2},
\eea
where
\be
\langle k,N-1|^{(0)}\p_{k}|{\rm FS}\rangle = \frac{1}{L^{1/2}} +F_1+F_2+F_3 + O(\hat{V^4}). 
\ee \ew

Let us now consider each of the terms in Eqs.~(\ref{f1}) - (\ref{f3}) on the basis of allowed intermediate states. We observe that the only admissible states $|\alpha\rangle$ in \ceq{f1} are ones containing a net zero momentum pair of particle-hole pairs and a hole of momentum $k$. This is because the third matrix element in \ceq{f1} will lead to states $|\beta\rangle$ with two net-zero momentum particle-hole pairs, while the second matrix element can indeed admit two kinds of states. However, the matrix element of the annihilation operator will only connect $|\alpha\rangle$ and $|\beta\rangle$ that have identical pairs of net zero momentum particle-hole pairs, since the annihilation operator can then create the additional hole at $k$ in $\beta$. Thus we may write the total contribution due to $F_1$ as
\bea
\label{f1discrete}
F_1=\hspace{7.5 cm}\nonumber\\
\sum\limits_{p\neq k, p\geq -k_F}^{k_F}\sum\limits_{q>k_F-p}^{\infty}\sum\limits_{p'\neq k,p, p'\geq -k_F}^{{\rm min}[k_F,q-k_F]}\frac{2[V(q)-V(p-p'+q)]^2}{4L^{5/2}(\Delta E)^2},\nonumber\\
\Delta E = -\frac{q}{m}(q+(p-p')).\ \ \ \ \ \ \ \ \ 
\eea

In \ceq{f1discrete} we enumerate all possible ways to generate an intermediate state with two net zero momentum particle-hole pairs. We can understand the limits on the sums in the following way: the first hole can be placed anywhere in the filled Fermi sea (except at the pre-existing hole of momentum $k$). With this hole in place, the momentum transfer to the corresponding particle (total momentum of $p+q$) needs to be large enough that the particle is created ``outside'' the filled Fermi sea. We first consider a positive momentum transfer and have that the transfer $q$ must be larger than $k_F - p$. This sets the lower limit on the second sum. Finally we consider the second particle-hole pair. For the hole of momentum $p'$ to lie within the Fermi sea and simultaneously have its corresponding particle of momentum $p'-q$ outside the sea, we note that $p'$ can have at most a momentum of $q-k_F$ provided that $q-k_F \leq k_F$. If $q$ exceeds this latter condition then $p'$ may lie anywhere in the sea as long as it is not equal to $p$ or $k$. If we then consider how the intermediate state corresponding to some allowed choice of $p,p',q$ can be created with a negative momentum transfer we find that setting $q \to p'-p-q$ also generates the same state. Moreover we may also relabel $p\to p', p' \to p$ and repeat the same argument above, leading to an additional factor of 2 multiplying the sums. The relative signs of these terms is fixed by the canonical anti-commutation relations of the fermionic operators. Furthermore the only way to connect states $\alpha$ and $\beta$ in \ceq{f1} with the annihilation operator matrix product is if the particle-hole pairs in the two states are identical, leading to the appearance of the square in \ceq{f1discrete}.

A very similar argument to the one above allows us to write down an expression for $F_2$. The key difference is there is no restriction disallowing the holes from having a momentum $k$ since the intermediate states are not involved in the annihilation operator matrix element. Thus we have

\bea
\label{f2discrete}
F_2=\hspace{7.5 cm}\nonumber\\
-\sum_{ p\geq -k_F}^{k_F}\sum_{q>k_F-p}^{\infty}\sum_{\substack{ p'\neq p,\\p'\geq -k_F}}^{{\rm min}[k_F,q-k_F]}\frac{2[V(q)-V(p-p'+q)]^2}{8L^{5/2}(\Delta E)^2},\nonumber\\
\Delta E = -\frac{q}{m}(q+(p-p')).\ \ \ \ \ \ \ \
\eea

Lastly, for $F_3$ we have two groups of terms. On the one hand akin to Eq.~(\ref{f1discrete}) we may have intermediate states with two particle-hole pairs. Here too there is a restriction on where the holes can be since there is a pre-existing hole of momentum $k$. On the other hand, we may also have states where the hole of momentum $k$ has been `moved', followed by the creation of an additional particle-hole pair. 

\bea
\label{f3discrete}
& &F_3 = \hspace{7.5 cm}\nonumber\\
 &-&\sum_{\substack{p\neq k,\\ p\geq -k_F}}^{k_F}\sum_{q>k_F-p}^{\infty}\sum_{\substack{p'\neq k,p,\\ p'\geq -k_F}}^{{\rm min}[k_F,q-k_F]}\frac{2[V(q)-V(p-p'+q)]^2}{8L^{5/2}(\Delta E)^2} \nonumber\\
&-&\sum_{p=-k_F}^{p<k}\sum_{p'=-k_F}^{-k_F-p+k}\frac{2[V(k-p)-V(k-p')]^2}{8L^{5/2}(\Delta \tilde{E})^2} \nonumber\\
&-& \sum_{p>k}^{k_F}\sum_{p'>k_F-p+k}^{k_F} \frac{2[V(k-p)-V(k-p')]^2}{8L^{5/2}(\Delta \tilde{E})^2},     \nonumber\\
&&\Delta E =-\frac{q}{m}(q+(p-p')), \nonumber\\
&&\Delta \tilde{E} = \frac{k}{m}(p+p'-k) - \frac{pp'}{m}.
\eea

Thus upon summing these contributions we are left with the following terms:
\bea
\sum_i F_i&=&-(T_{\rm 4ex} + T_{\rm 3ex}^{(1)} + T_{\rm 3ex}^{(2)}), \nonumber\\
 T_{\rm 3ex}^{(1)} &=&\sum_{p=-k_F}^{p<k}\sum_{p'=-k_F}^{-k_F-p+k}\frac{[V(k-p)-V(k-p')]^2}{4L^{5/2}(\Delta \tilde{E})^2},\nonumber\\
 T_{\rm 3ex}^{(2)}&=& \sum_{p>k}^{k_F}\sum_{p'>k_F-p+k}^{k_F} \frac{[V(k-p)-V(k-p')]^2}{4L^{5/2}(\Delta \tilde{E})^2},\nonumber\\
 T_{\rm 4ex} &=& \sum_{q>k_F-k}^{\infty} \sum_{\substack{p'\neq k,\\p'\geq-k_F}}^{{\rm min}[q-k_F,k_F]}\frac{[V(q)-V(k-p'+q)]^2}{4L^{5/2}(\Delta E)^2},\nonumber\\
\Delta E &=& -\frac{q}{m}(q+(k-p')),\nonumber\\
 \Delta \tilde{E} &=& \frac{k}{m}(p+p'-k) - \frac{pp'}{m}.
\eea
The notation above is suggestive of the fact that we are considering contributions to the form factor due to intermediate states with 4 excitations (2 particle-hole pairs) and 3 excitations (a particle-hole pair and an unpaired hole) separately.

Let us consider first:

\bea
\label{t13ex}
T^{(1)}_{{\rm 3ex}}=\frac{m^2}{4L^{5/2}}\sum_{p=-k_F}^{p<k}\sum_{p'=-k_F}^{p'<-k_F+k-p}\frac{[V(k-p)-V(k-p')]^2}{(k-p)^2(p'-k)^2}  \ \ \ \ \ \ \  \ \ \ \ \ \ \ \  \nonumber\\
=\frac{m^2}{2L^{1/2}}\int\limits_{\frac{-k_F+k}{2}}^{k-\Delta}\frac{dp}{2\pi}\int\limits_{-k_F}^{-k_F+k-p}\frac{dp'}{2\pi}\frac{[V(k-p)-V(k-p')]^2}{(k-p)^2(p'-k)^2}  \ \ \ \ \ \ \  \ \ \ \ \ \ \ \ \nonumber\\
+ \frac{m^2}{2L^{5/2}}\sum_{p=k-\Delta}^{p=k-2\pi/L}\sum_{p'=-k_F}^{p'=-k_F+k-p-2\pi/L}\frac{[V(k-p)-V(k-p')]^2}{(k-p)^2(p'-k)^2}  \ \ \ \ \ \ \  \ \ \ \ \ \ \ \nonumber\\
+\frac{m^2}{4L^{1/2}}\int\limits_{-k_F}^{\frac{-k_F+k}{2}}\frac{dp}{2\pi}\int\limits_{-k_F}^{\frac{-k_F+k}{2}}\frac{dp'}{2\pi}\frac{[V(k-p)-V(k-p')]^2}{(k-p)^2(p'-k)^2}.\ \ \ \ \ \ \  \ \ \ \ \ \ \ \ \nonumber
\eea

\begin{figure}
\includegraphics[width=8.5 cm,height=6.5cm]{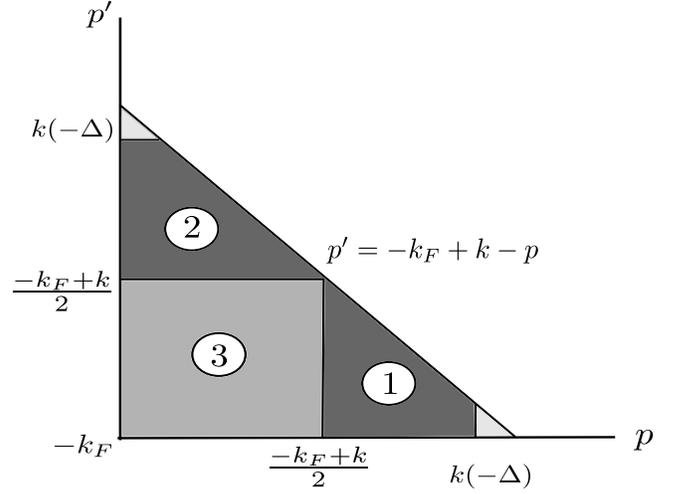}
\caption{\label{Fig_int1} The various regions of integration corresponding to summing up the contribution to the form factor from states that have two holes and one particle excitation.}
\end{figure}

Starting from the initial sum, we have separated the contributions from three different regions over which we need to sum  the right side of the above equation. We have taken the continuum limit to express the sums as integrals in these regions, however, we have separated the contributions in regions (1) and (2) when one of the holes is within the region of momentum space, $(k-\Delta,k)$ (corresponding to the lightly shaded triangles in Fig.~(\ref{Fig_int1})), and maintained these contributions as discrete sums. This is because the energy denominator will become badly behaved in this region and upon integration yield a logarithm of $L$; however we require more than logarithmic accuracy since such corrections may enter the prefactor. 

While our expressions are formally correct for a fixed $\Delta$. We wish to demonstrate that the limit as $\Delta \to 0$ of the above expression exists and subsequently obtain an analytic expression for it. In order to do this we will approximate the integral when the range of $p$ nears $k-\Delta$ and obtain the leading order $\Delta$ dependent contribution. We will then carry out the discrete sums and obtain the leading order $\Delta$ dependent contribution as well as any non-vanishing constants. We will show the existence of the $\Delta \to 0$ limit by demonstrating that all divergent $\Delta$ dependent contributions vanish. 

After approximating the behavior of the divergent integral in \ceq{t13ex} at the upper bound we are left with
\bw
\bea
\frac{m^2}{16\pi^2k^2L^{1/2}}\mathcal{P}_+\int\limits_{\frac{1-k_F/k}{2}}^{1}dx\int\limits_{-k_F/k}^{1-k_F/k-x}dy \frac{[V(k(1-x))-V(k(1-y))]^2}{(1-x)^2(y-1)^2}-\frac{m^2[V(0)-V(k+k_F)]^2}{4\pi^2L^{1/2}(k+k_F)^2}\left(\log(\Delta)\right),
\eea
\ew
where the notation and meaning of the special principal value integration denoted by $\mathcal{P}_+$ is defined in the appendix, see Eqs.(\ref{pints})-(\ref{avoidconfusion}).

Meanwhile we may evaluate the sum in \ceq{t13ex} explicitly using the identities and expansions of the poly-gamma functions, Eqs~(\ref{sum_1})-(\ref{asymptotics}):

\bw
\bea
& & \frac{m^2}{2L^{5/2}}\sum_{p=k-\Delta}^{p=k-2\pi/L}\sum_{p'=-k_F}^{p'=-k_F+k-p-2\pi/L}\frac{[V(k-p)-V(k-p')]^2}{(k-p)^2(p'-k)^2}\nonumber\\
&=&\frac{m^2[V(0)-V(k+k_F)]^2L^{3/2}}{32\pi^4((N-1)/2+n_k)^2}\left(\log\left(\frac{2\pi \delta}{L}\right) - \log\left(\frac{2\pi}{L}\right) + \gamma_E\right)\nonumber\\
&=& \frac{m^2[V(0)-V(k+k_F)]^2}{8\pi^2 L^{1/2}(k_F+k)^2}\left(\log(\Delta) - \log\left(\frac{2\pi}{L}\right) + \gamma_E\right),
  \eea
  \ew
where $\gamma_E \approx 0.5772$ is the Euler-Mascheroni constant.

  We note how the logarithmic divergence appears with opposite sign in the finite sum and the boundary of the integral and thus drops out of the final answer. We therefore obtain an expression for the term
\bw
  \bea
  T^{(1)}_{\rm 3ex} &=& \frac{m^2}{4L^{1/2}}\mathcal{P}_+\int\limits_{\frac{1-k_F/k}{2}}^{1}\frac{dx}{2\pi}\int\limits_{-k_F/k}^{-k_F/k+1-x}\frac{dy}{2\pi}\frac{[V(k(1-x))-V(k(1-y))]^2}{k^2(1-x)^2(y-1)^2}\nonumber\\
  &+&\frac{m^2}{4L^{1/2}}\int\limits_{-k_F}^{\frac{-k_F+k}{2}}\frac{dp}{2\pi}\int\limits_{-k_F}^{\frac{-k_F+k}{2}}\frac{dp'}{2\pi}\frac{[V(k-p)-V(k-p')]^2}{(k-p)^2(p'-k)^2} - \frac{m^2[V(0)-V(k+k_F)]^2}{8\pi^2L^{1/2}(k+k_F)^2}\left(\log\left(\frac{2\pi}{L}\right) - \gamma_E\right).
\eea
\ew

A similar calculation yields for the second term summing contributions from intermediate states with three excitations
\bw
\bea
T^{(2)}_{\rm 3ex} &=& \frac{m^2}{4L^{1/2}}\mathcal{P}_-\int\limits^{\frac{1+k_F/k}{2}}_{1}\frac{dx}{2\pi}\int\limits^{k_F/k}_{k_F/k+1-x}\frac{dy}{2\pi}\frac{[V(k(x-1))-V(k(y-1))]^2}{k^2(1-x)^2(y-1)^2}\nonumber\\
&+&\frac{m^2}{4L^{1/2}}\int\limits^{k_F}_{\frac{k_F+k}{2}}\frac{dp}{2\pi}\int\limits^{k_F}_{\frac{k_F+k}{2}}\frac{dp'}{2\pi}\frac{[V(p-k)-V(p'-k)]^2}{(k-p)^2(p'-k)^2}
-\frac{m^2[V(0)-V(k_F-k)]^2}{8\pi^2L^{1/2}(k_F-k)^2}\left(\log\left(\frac{2\pi}{L}\right) - \gamma_E\right).
\eea
\ew

The sums from the contribution of intermediate states with two pairs of particle-hole excitations are well behaved and can be immediately interpreted as integrals:
\bw
\bea
T_{\rm 4ex} &=& \frac{m^2}{4L^{5/2}}\sum_{q>k_F-k}^{\infty}\sum_{p' >-k_F}^{{\rm min}[-k_F+q,k_F]}\frac{[V(q)-V(k-p'+q)]^2}{q^2(k-p'+q)^2}\nonumber\\
&=& \frac{m^2}{16\pi^2L^{1/2}}\left[\int_{k_F-k}^{2k_F}dq\int_{-k_F}^{-k_F+q}dp'\frac{[V(q)-V(k-p'+q)]^2}{q^2(q+k-p')^2} + \int_{2k_F}^{\infty}dq\int_{-k_F}^{-k_F}dp'\frac{[V(q)-V(k-p'+q)]^2}{q^2(q+k-p')^2}\right]
\eea
\ew

Thus we obtain the full contribution to the form factor up to order $\V^2$:
\bw
\bea\label{ff_a0_answer}
& &\langle k,N-1|\p(0)|N\rangle\nonumber\\
 &=& \frac{1}{2L^{1/2}}\bigg[2 - \frac{m^2}{4\pi^2}\bigg\{\mathcal{P}_+\int\limits_{\frac{1-k_F/k}{2}}^{1}dx\int\limits_{-k_F/k}^{-k_F/k+1-x}dy\frac{[V(k(1-x))-V(k(1-y))]^2}{k^2(1-x)^2(y-1)^2}\nonumber\\
 & &+ \mathcal{P}_-\int\limits^{\frac{1+k_F/k}{2}}_{1}dx\int\limits^{k_F/k}_{k_F/k+1-x}dy\frac{[V(k(x-1))-V(k(y-1))]^2}{k^2(1-x)^2(y-1)^2}\nonumber\\
& &+ \int\limits_{-k_F}^{\frac{-k_F+k}{2}}dp\int\limits_{-k_F}^{\frac{-k_F+k}{2}}dp'\frac{[V(k-p)-V(k-p')]^2}{2(k-p)^2(p'-k)^2}+\int\limits^{k_F}_{\frac{k_F+k}{2}}dp\int\limits^{k_F}_{\frac{k_F+k}{2}}dp'\frac{[V(p-k)-V(p'-k)]^2}{2(k-p)^2(p'-k)^2}\nonumber\\
& &+\int_{k_F-k}^{2k_F}dq\int_{-k_F}^{-k_F+q}dp'\frac{[V(q)-V(k-p'+q)]^2}{2q^2(q+k-p')^2} + \int_{2k_F}^{\infty}dq\int_{-k_F}^{-k_F}dp'\frac{[V(q)-V(k-p'+q)]^2}{2q^2(q+k-p')^2}\nonumber\\
& &+\left(\frac{[V(0)-V(k_F+k)]^2}{(k_F+k)^2} + \frac{[V(0)-V(k_F-k)]^2}{(k_F-k)^2}\right)\gamma_E\bigg\} \nonumber\\
& &+\frac{m^2}{4\pi^2}\left(\frac{[V(0)-V(k_F+k)]^2}{(k_F+k)^2} + \frac{[V(0)-V(k_F-k)]^2}{(k_F-k)^2}\right)\log\left(\frac{2\pi}{L}\right)\bigg].\nonumber\\
\eea
\ew

Comparing \ceq{ff_a0_answer} with  the field theory prediction, we obtain the leading correction to the prefactor $A_{0,-}(k)$:
\bw
 \bea
 A_{0,-}(k) &=& \frac{1}{2\pi}  - \frac{m^2}{8\pi^3}\bigg\{\mathcal{P}_+\int\limits_{\frac{1-k_F/k}{2}}^{1}dx\int\limits_{-k_F/k}^{-k_F/k+1-x}dy\frac{[V(k(1-x))-V(k(1-y))]^2}{k^2(1-x)^2(y-1)^2}\nonumber\\
 & &+ \mathcal{P}_-\int\limits^{\frac{1+k_F/k}{2}}_{1}dx\int\limits^{k_F/k}_{k_F/k+1-x}dy\frac{[V(k(x-1))-V(k(y-1))]^2}{k^2(1-x)^2(y-1)^2}\nonumber\\
& &+ \int\limits_{-k_F}^{\frac{-k_F+k}{2}}dp\int\limits_{-k_F}^{\frac{-k_F+k}{2}}dp'\frac{[V(k-p)-V(k-p')]^2}{2(k-p)^2(p'-k)^2}+\int\limits^{k_F}_{\frac{k_F+k}{2}}dp\int\limits^{k_F}_{\frac{k_F+k}{2}}dp'\frac{[V(p-k)-V(p'-k)]^2}{2(k-p)^2(p'-k)^2}\nonumber\\
& &+\int_{k_F-k}^{2k_F}dq\int_{-k_F}^{-k_F+q}dp'\frac{[V(q)-V(k-p'+q)]^2}{2q^2(q+k-p')^2} + \int_{2k_F}^{\infty}dq\int_{-k_F}^{-k_F}dp'\frac{[V(q)-V(k-p'+q)]^2}{2q^2(q+k-p')^2}\nonumber\\
& &+\left(\frac{[V(0)-V(k_F+k)]^2}{(k_F+k)^2} + \frac{[V(0)-V(k_F-k)]^2}{(k_F-k)^2}\right)\gamma_E\bigg\}.
\eea
\end{widetext}
\newpage

\section{Summary}
We present below a few results for the nonuniversal prefactors of correlation functions and dynamic response functions (see Eqs.~(\ref{Amdef})-(\ref{Cmdef}) and Eq.~(\ref{Aint})).  The results of the perturbative calculations described in the previous section produce the following leading order results for the prefactors (see Eqs.~(\ref{Aint}),(\ref{A_0_p_def}), (\ref{A_1_m_def}), (\ref{Amdef}), and (\ref{Cmdef}) respectively):
\bea
A_{0,+}(k) &=& \frac{M^2(V(k_F+k)-V(k_F-k))^2}{8\pi^3(k_F^2-k^2)^2},\label{a0plus}\\
A_{1,-}(k) &=& \frac{M^2(V(2k_F)-V(k_F-k))^2}{32\pi^3k_F^2(k-k_F)^2},\\
A_{1,+}(k) &=&\frac{M^2 V'(k-k_F)^2}{8\pi^3 (k-k_F)^4},\\ 
C_1 &=& \frac{M^2\rho_0 V'(2k_F)^2}{64\pi^7},\ \ \ \ \\
A_2 &=&  \frac{M^2\rho_0(V'(2k_F)-k_FV''(2k_F))^2}{32\pi^{10}}. \label{a2}
\eea

We see that all the prefactors above vanish when $V(k)$ is constant, satisfying the expectation that a contact interaction $V(r)\propto\delta(r)$ between the spinless fermions should not affect any observables. We were also able to reproduce the known values of the prefactors $C_1, A_2$ \cite{CS_results} for the Calogero-Sutherland model, with our perturbative fermionic results, Eqs.~(22), (23). We note that the results above 
would be hard to obtain using the conventional infinite size diagrammatic technique due to logarithmic divergences~\cite{Pustilnik2006Fermions,Khodas2006Fermions},
but the consideration of finite size scaling allows us to obtain these answers using essentially only undergraduate quantum mechanics tools.
To lowest non-vanishing order, we see that the prefactors $A_{0,+}(k),A_{1,\pm}(k)$ are $\propto V^2.$ In general, to lowest non-vanishing order $A_{n\geq1,\pm}(k)$ will be $\propto V^{2n}$,
although the complexity of the expressions grows. To obtain corrections beyond leading order to $A_{n,\pm}(k)$ (and the first non-trivial correction to $A_{0,-}(k), A_1$ and $C_0$), one needs to carefully separate terms in the perturbation series that diverge as powers of $\log(L)$ from the relevant correction to the prefactors,
which we have demonstrated in Section~IIIF above.

\begin{figure}
\includegraphics[width=8.5cm,height=6 cm]{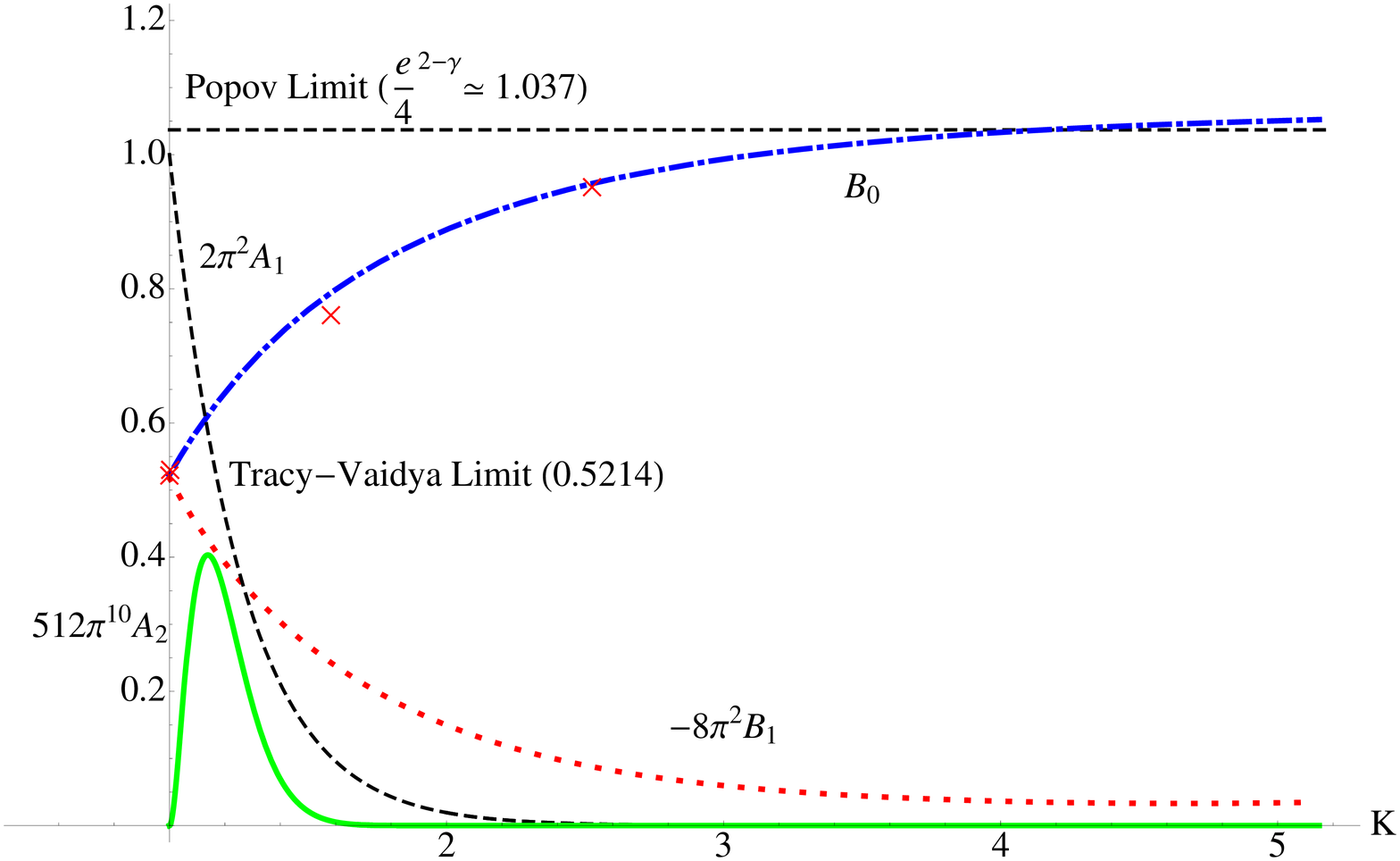}
\caption{\label{FigLL}(Color online) Non-perturbative results for the Lieb-Liniger model of 1D bosons:  $ 2\pi^{2}A_1$ (dashed black), $512\pi^{10} A_2$ (solid green) and $B_0$ (dot-dashed blue), $-8\pi^2B_1$ (dotted red) as functions of the Luttinger liquid parameter $K$. In the limit of strong interaction ($K \to 1$) our expressions for $B_0$ and $B_1$ agree with the known values~\cite{K1}, while $A_1 \to 1/2\pi^2, A_2 \to 0$ are in accordance with the density correlator of the free Fermi gas. We also match $B_0$ in the weakly interacting regime ($K \gg 1$) to Popov's result (dashed line) \cite{Popov_prefactor}, and  show some numerical results (crosses)~\cite{Numerical_prefactors}.
 }
\end{figure}

For the Lieb-Liniger model, we obtain  analytic expressions for the prefactors of correlation functions non-perturbatively. We relate them to the exactly known form factors~\cite{Slavnov}  which have been used recently for numerical evaluation of dynamic response functions~\cite{Caux}. In Fig.~\ref{FigLL} we show results for a few prefactors for different interaction strengths (details will be presented elsewhere~\cite{long}), and some known limits are also plotted for comparison. As a further check we have also successfully reproduced the perturbative fermionic result for the prefactor $A_2$ [see Eq.~(\ref{a2})] of the Cheon-Shigehara model~\cite{CSh}, which is  dual to the Lieb-Liniger model and has the same density correlations. It should be noted that recently the correlation functions for the Lieb-Liniger model have been treated without field theoretical considerations but rather by directly working with the microscopic details of the theory (see Ref.~\cite{Kozlowski}), and the results of this work coincide with ours.

To summarize, we developed a general approach to calculating nonuniversal prefactors in static and dynamic correlation functions of 1D quantum liquids,
by relating them to the finite-size scaling of the matrix elements of the corresponding operators. To find a given prefactor, only a single matrix element (form factor) between the lowest energy states needs to be evaluated, see Eqs.~(\ref{fermion_scaling})-(\ref{density_scaling}), (\ref{Ascaling}). Moreover, the method does not rely on the integrability of a model. 
To demonstrate our approach, we calculated some prefactors in static and dynamic correlation functions for weakly interacting spinless fermions with an arbitrary pair interaction potential, see Eqs.~(\ref{a0plus})-(\ref{a2}).

 Methods of Section II applied to an integrable model, allow one to obtain exact expressions for prefactors of the response functions, at arbitrary interaction strength. The results of the perturbative calculation detailed in Section III agree, in the proper limit, with the results~\cite{long} obtained by  the methods of Section II for the exactly solvable Lieb-Liniger model, see Fig. \ref{FigLL}.

\section{Acknowledgements}

This work was supported by the NSF DMR Grant No. 0906498, the Texas NHARP Grant No. 01889, the FOM foundation and the Alfred P. Sloan Foundation.

\begin{appendix}
\section{Multiplet Summation Rule}

 \begin{figure}
\includegraphics[width=8.5 cm,height=3.8cm]{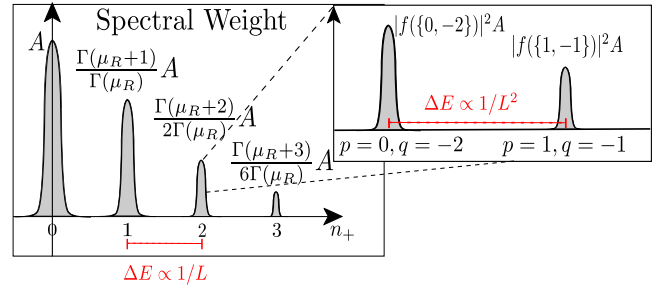}
\caption{\label{multiplet_splitting}(Color online) Schematic showing how spectral weights are distributed amongst multiplet of states which are degenerate within $\propto 1/L$ accuracy.
 In the inset, we illustrate how spectral weights are distributed within a multiplet for $n_+=2,$ once the degeneracy is lifted by $\propto 1/L^2$ corrections. }
\end{figure}

Let us now provide an illustration of the multiplet summation rule, Eq.~(\ref{mrule}) of the article, for a few simple cases. The sum rule is given by
\be
\label{sum_rule}
\sum_{\sum p_i-q_i = n_{+}} | f(\{p_i\}; \{q_i\})|^2 =  C(n_{+}, \mu_{F,R}),
\ee
where
\bea
C(n_r,\mu_{F,R}) = \frac{\Gamma(\mu_{F,R} + n_{+})}{\Gamma(\mu_{F,R})\Gamma(n_{+} + 1)},\nonumber\\
\eea
and $f(\{p_i,q_i\})$ is defined in \ceq{f_def}. The rules for $n_-$ are analogous.

\subsection{$n_+ = 1$}
For $n_+ = 1$, the only state which contributes to the multiplet is $p = 0, q = -1$. Thus the left hand side of \ceq{sum_rule} is given by
\bea
|f(\{0,-1\})|^2 &=& |f^+(0)\times f^-(-1)|^2 \nonumber\\
&=&\bigg|\left(\frac{(-\sqrt{\mu_{F,R}})(-\sqrt{\mu_{F,R}}-1)...}{(-\sqrt{\mu_{F,R}}-1)...}\right)\nonumber\\
&\times&\left(\frac{(\sqrt{\mu_{F,R}}-1)(\sqrt{\mu_{F,R}} - 2)...}{(\sqrt{\mu_{F,R}}-1)(\sqrt{\mu_{F,R}} - 2)...}\right)\bigg|^2\nonumber\\
&=& \mu_{F,R}.
\eea

The right hand side of \ceq{sum_rule} is given by
\bea
C(n_+=1, \mu_{F,R}) &=& \left(\frac{(\mu_{F,R})(\mu_{F,R} - 1)...}{(\mu_{F,R} - 1)...}\right) = \mu_{F,R}\nonumber\\
&=& |f(\{0,-1\})|^2 .
 \eea

\subsection{$n_+ = 2$}

For $n_+ =2$, there are two states: $p=1, q= -1$, and $p=0, q= -2$. Consequently the left hand side of \ceq{sum_rule} is given by
\bea
& &|f(+1;-1)|^2 + |f(0,-2)|^2 = \nonumber\\
& & \left|\frac{1}{2}f^+(1)f^-(-1)\right|^2 +  \left|\frac{1}{2}f^+(0)f^-(-2)\right|^2\nonumber\\
&=& \frac{\mu_{F,R}(\sqrt{\mu_{F,R}} - 1)^2}{4} + \frac{\mu_{F,R}(\sqrt{\mu_{F,R}} + 1)^2}{4}\nonumber\\
&=& \frac{\mu_{F,R}(\mu_{F,R} + 1)}{2}.
\eea

 And the right hand side of \ceq{sum_rule} is given by
\bea
& &C(n_+=2,\mu_{F,R}) = \left(\frac{(\mu_{F,R}+1)(\mu_{F,R})(\mu_{F,R} - 1)...}{\Gamma(3)(\mu_{F,R}  - 1)... }\right)\nonumber\\
&=& \frac{\mu_{F,R}(\mu_{F,R}+1)}{2} = |f(+1;-1)|^2 + |f(0,-2)|^2 .
\eea

\subsection{$n_+ = 3$}

There are three states in the multiplet $n_+=3$: $p=2, q=-1, p=1, q=-2$ and $p=0, q=-3.$ The left hand side of \ceq{sum_rule} is given by
\bea
& &|f(+2;-1)|^2 + |f(+1;-2)|^2 + |f(0;-3)|^2\nonumber\\
&=&\frac{1}{36}\mu_R(2-\sqrt{\mu_R})^2(1-\sqrt{\mu_R})^2 \nonumber\\
& &+ \frac{1}{9}\mu_R(\sqrt{1-\mu_R})^2(1+\sqrt{\mu_R})^2\nonumber\\
& &+\frac{1}{36}\mu_R(2+\sqrt{\mu_R})^2(1+\sqrt{\mu_R})^2\nonumber\\
 &=& \frac{\Gamma(\mu_R + 3)}{6\Gamma(\mu_R)} = C(n_+=3,\mu_R).
\eea

Thus once again the summation rule holds.

\section{Identities involving Poly-Gamma functions}
The poly-gamma function is defined as $\psi^{(k)}(z) = \frac{d^k}{dx^k}\log(\Gamma(x))|_{x=z}$.
We have the following summation identity for $\psi^{(n)}(z)$:
\bea
\label{sum_1}
\sum_{k=0}^{N-1}\frac{1}{(k+a)^{n+1}} = \frac{(-1)^n}{n!}(\psi^{(n)}(N+a) - \psi^{(n)}(a)).
\eea
We have a few explicit values of the polygamma functions:
\bea
\psi^{(0)}(1) = -\gamma_E,\nonumber\\
\psi^{(1)}(1) = \frac{\pi^2}{6},
\eea
where $\gamma_E \approx 0.5772...$ is the Euler-Mascheroni constant. 

We also have the following asymptotic expansions:
\bea
\label{asymptotics}
\lim_{z\to \infty} \psi^{(0)}(z) = \log(z) -\frac{1}{2z} + O\left(\frac{1}{z^2}\right),\nonumber\\
\lim_{z\to \infty} \psi^{(1)}(z) = \frac{1}{z} +\frac{1}{2z^2} + O\left(\frac{1}{z^3}\right).
\eea

\section{Principal Value Integrals: Interior and Edge Singularities}

We define three distinct types of principal value integrals as follows,
\bea
\label{pints}
& &P \int_{a}^{b} dx \frac{f(x)}{x-c} = \lim_{\delta \rightarrow 0} \left( \int_{a}^{c-\delta} dx \frac{f(x)}{x-c} + \int_{c + \delta}^{b} dx \frac{f(x)}{x-c}\right),\nonumber\\
& &P_{-} \int_{a}^{b} dx \frac{f(x)}{x-a} = \lim_{\delta \rightarrow 0} \left( \int_{a+\delta}^{b} \frac{f(x)}{x-a} +f(a)\log(\delta)\right), \nonumber \\
& &P_{+} \int_{a}^{b} dx \frac{f(x)}{x-b}= \lim_{\delta \rightarrow 0} \left(\int_{a}^{b-\delta} \frac{f(x)}{x-b} - f(b)\log(\delta)\right),\nonumber\\
& &P_{\pm} \int_{a}^{b} dx \frac{f(x)(b-a)}{(x-a)(x-b)} = P_+ \int_{a}^{b} \frac{f(x)}{x-b} - P_- \int_{a}^{b} \frac{f(x)}{x-a}. \nonumber\\
\eea

Note also that when we have a multiple integral where all but one of the integrals is analytic and do not need to be evaluated in the special sense above, we will use the notation
\bea
\label{avoidconfusion}
P_{\pm} \int_{a}^{b} dx \int_{c}^{d}dy \frac{f(x)f(y)}{x-b(a)},
\eea
where we mean that we can freely carry out the inner integration, but the final integration needs to be performed as in \ceq{pints}.
\end{appendix}


\begin{thebibliography}{19}

\bibitem{EL} K.B. Efetov and A.I. Larkin, Sov. Phys. JETP {\bf 42}, 390
(1975).
\bibitem{Haldane} F.D.M. Haldane, Phys. Rev. Lett. {\bf 47}, 1840 (1981); J. Phys. C: Solid State Phys. {\bf 14}, 2585 (1981).
\bibitem{Caza04} M.A. Cazalilla, J. Phys. B  {\bf 37}, S1 (2004).
\bibitem{Gbook}T. Giamarchi, {\it Quantum Physics in One Dimension} (Clarendon, Oxford, 2004).
\bibitem{GNT} A. Gogolin, A. Nersesyan, and A. Tsvelik, {\it Bosonization and Strongly Correlated Systems} (Cambridge University Press, 1999).



\bibitem{Popov_prefactor}V. N. Popov, JETP Letters {\bf 31}, 526 (1980).
\bibitem{K1}H.G. Vaidya and C.A. Tracy, Phys. Rev. Lett. {\bf 43}, 1540 (1979); J. Math. Phys. 20, 2291 (1979); M. Jimbo, T. Miwa, Y. Mori, and M. Sato, Physica (Utrecht) {\bf 1D}, 80
(1980); D. M. Gangardt, J. Phys. A {\bf 37}, 9335 (2004).
\bibitem{CS_results} D. M. Gangardt and A. Kamenev, Nucl. Phys. B, {\bf 610}, 578 (2001).
\bibitem{lukyanov03}S.~Lukyanov and V.~Terras, Nucl.Phys. {\bf B654,}  323 (2003).
\bibitem{Kitanine_review} N. Kitanine, K. K. Kozlowski, J. M. Maillet, N. A. Slavnov, and V. Terras  J.~Stat. Mech. (2009)
P04003.

\bibitem{Pustilnik2006Fermions}M. Pustilnik, M. Khodas, A. Kamenev, and L. I. Glazman, Phys. Rev. Lett. {\bf 96}, 196405 (2006).
\bibitem{Pustilnik_solo_PRL}M. Pustilnik, Phys. Rev. Lett. {\bf 97}, 036404 (2006).
\bibitem{Khodas2006Fermions}M. Khodas, M. Pustilnik, A. Kamenev, and L. I. Glazman, Phys. Rev. B {\bf 76}, 155402
(2007); Phys. Rev. Lett. {\bf 99}, 110405 (2007).
\bibitem{PRL_08}  A. Imambekov and  L. I. Glazman, Phys. Rev. Lett. {\bf 100}, 206805 (2008).
\bibitem{Zvonarev}M. B. Zvonarev, V. V. Cheianov, and T. Giamarchi, Phys. Rev. Lett. {\bf 103}, 110401 (2009)
\bibitem{universal}A.~Imambekov and L.I.~Glazman, Science {\bf 323}, 228 (2009).
\bibitem{PRL_09}A.~Imambekov and L.I.~Glazman, Phys. Rev. Lett. {\bf 102}, 126405 (2009).
\bibitem{PRL_10}T.L.~Schmidt, A.~Imambekov, and L.I.~Glazman, Phys. Rev. Lett. {\bf 104}, 116403 (2010);
Phys. Rev. B {\bf 82}, 245104 (2010); R. G. Pereira and E. Sela, Phys. Rev. B {\bf 82}, 115324 (2010); F. H. L. Essler Phys. Rev. B {\bf 81}, 205120 (2010); A. Kamenev and L. I. Glazman, Phys. Rev. A {\bf 80}, 011603 (2009); M. Khodas, A. Kamenev, and L. I. Glazman, Phys. Rev. A {\bf 78}, 053630 (2008)
\bibitem{XXZ}V.V. Cheianov and  M. Pustilnik, Phys. Rev. Lett. {\bf 100}, 126403 (2008); R.G. Pereira, S. R. White, and I. Affleck, {\it ibid.} {\bf 100}, 027206 (2008);
Phys. Rev. B {\bf 79}, 165113 (2009).



\bibitem{LL}E.H. Lieb and W. Liniger, Phys. Rev. {\bf 130}, 1605
(1963); E.H. Lieb, {\it ibid}. {\bf 130}, 1616 (1963).

\bibitem{Korepin} V.E. Korepin, N.M. Bogoliubov, and A.G. Izergin, {\it Quantum Inverse Scattering Method and Correlation Functions}
(Cambridge University Press, Cambridge, 1993).




\bibitem{dweiss} T. Kinoshita, T. Wenger, and D. S. Weiss, Science {\bf 305}, 1125
(2004); A. H. van Amerongen, J. J. P. van Es, P. Wicke, K. V. Kheruntsyan, and N. J. van Druten, Phys. Rev. Lett. {\bf 100}, 090402 (2008).  

\bibitem{interference}A. Polkovnikov, E. Altman, and  E. Demler, PNAS {\bf 103}, 6125 (2006);
A. Imambekov, V. Gritsev, and E. Demler, Phys. Rev. A {\bf 77},
063606 (2008); S. Hofferberth, I. Lesanovsky, T. Schumm, A. Imambekov, V. Gritsev, E. Demler, and J. Schmiedmaye, Nature Phys. {\bf 4}, 489
(2008).

\bibitem{exp2} T. Donner,  S. Ritter, T. Bourdel, A. \"Ottl, M. K\"ohl, and T. Esslinger, Science {\bf 315}, 1556 (2007).

\bibitem{3decay}B. Laburthe Tolra, K. M. O' Hara, J. H. Huckans, W. D. Phillips, S. L. Rolston, and J. V. Porto, 
Phys. Rev. Lett. {\bf 92}, 190401 (2004); T. Kinoshita, T. Wenger, and D.~S. Weiss,  {\it ibid.} {\bf 95}, 190406 (2005).

\bibitem{BraggPE}S.B. Papp, J. M. Pino, R. J. Wild, S. Ronen, C. E. Wieman, D. S. Jin, and E. A. Cornell, Phys. Rev. Lett. {\bf 101}, 135301
(2008);  G. Veeravalli, E. Kuhnle, P. Dyke, and C. J. Vale, {\it ibid.} {\bf 101}, 250403
(2008); D. Cl\'{e}ment, N. Fabbri, L. Fallani, C. Fort, and M. Inguscio, {\it ibid.} {\bf 102}, 155301
(2009); P.T. Ernst, S\"oren G\"otze, Jasper S. Krauser, Karsten Pyka, Dirk-S\"oren L\"uhmann, Daniela Pfannkuche, and Klaus Sengstock, Nature Phys. {\bf 6}, 56 (2010);
J.T. Stewart, J.P. Gaebler, and D.S. Jin, Nature (London) {\bf
454}, 744 (2008).

\bibitem{AffleckCardy}H. W. J. Bl\"{o}te, J. L. Cardy, and M. P. Nightingale, Phys. Rev. Lett. {\bf 56}, 742 (1986); I. Affleck, {\it ibid.} {\bf 56}, 746 (1986).




\bibitem{Kitanine_JMP_09} N. Kitanine, K. K. Kozlowski, J. M. Maillet, N. A. Slavnov, and V. Terras, J. Math. Phys. {\bf 50}, 095209 (2009).
\bibitem{Kitanine} N. Kitanine, J. M. Maillet, and V. Terras, Nucl. Phys. B, {\bf 554}, 679 (1999); {\bf 567}, 554 (2000).
\bibitem{CSMexact}Y. Kuramoto and Y. Kato, {\it Dynamics of One-Dimensional Quantum Systems} (Cambridge University Press, 2009); Z.N.C. Ha, {\it Quantum Many-Body Systems in One Dimension} (World Scientific, 1996); Phys. Rev. Lett. {\bf 73}, 1574 (1994); {\bf 74}, 620(E)
(1995); Nucl. Phys. B {\bf 435}, 604 (1995).
\bibitem{Slavnov} N. A. Slavnov, Teor. Mat. Fiz. {\bf 79}, 232
(1989); {\bf 82}, 389 (1990); T. Kojima, V. E. Korepin, and N. A.
Slavnov, Commun. Math. Phys. {\bf 188}, 657 (1997).


\bibitem{Periera_PRL_06} R. G. Pereira, S. R. White, and I. Affleck, Phys. Rev. Lett. {\bf 100}, 027206 (2008); R. G. Pereira, J. Sirker, J.-S. Caux, R. Hagemans, J. M. Maillet, S. R. White, and  I. Affleck, J. Stat. Mech. (2007) P08022. 


\bibitem{BAW06} E. Bettelheim, A. G. Abanov, and P. Wiegmann, J. Phys. A {\bf 40, } F193 (2007).

\bibitem{AGDbook} A. A. Abrikosov, L. P. Gorkov, and I. E. Dzyaloshinskii, {\it Methods of Quantum Field Theory in Statistical Physics} (Dover Publications, 1975). 



\bibitem{muRLdef} $\mu_{n,\pm,R(L)}=\left[n\sqrt{K}-(+) \frac{(1\pm1)}{2\sqrt{K}}-\frac{\delta_{+(-)}(k_n)}{2\pi}\right]^2,\\
\mu^b_{n,\pm,R(L)}={\displaystyle \left[\frac{(2n+1)\sqrt{K}}{2}-(+) \frac{(1\pm1)}{2\sqrt{K}}-\frac{\delta_{+(-)}(k^*_n)}{2\pi}\right]^2},\\
\mu_{n,R(L)}={\displaystyle \left[\frac{(2n+1)\sqrt{K}}{2}+(-) \frac{1}{2\sqrt{K}}+\frac{\delta_{+(-)}(k^*_n)}{2\pi}\right]^2},$
in the notations of Ref.~\cite{PRL_09}.


\bibitem{Caux}J.-S. Caux and  P. Calabrese, Phys. Rev. A {\bf 74}, 031605(R)
(2006); J.-S. Caux, P. Calabrese, and  N.A. Slavnov, J.~Stat. Mech. (2007) P01008.



\bibitem{long} A. Shashi {\it et al.,}, arXiv:1010.2268v4.

\bibitem{CSh} T.~Cheon and T.~Shigehara, Phys. Rev. Lett. {\bf 82}, 2536 (1999);
Phys. Lett. A {\bf 243}, 111 (1998).

\bibitem{Kozlowski} K. K. Kozlowski, arXiv:1101.1626v2; K. K. Kozlowski and V. Terras, arXiv:1101.0844v1.


\bibitem{Numerical_prefactors}G. E. Astrakharchik and S. Giorgini, Phys. Rev. A {\bf 68}, 031602(R) (2003); J. Phys. B. {\bf 39} S1 (2006).




\end{thebibliography}
\end{document}